\documentclass[10pt,letterpaper]{article}
\usepackage[top=0.85in,left=2.75in,footskip=0.75in]{geometry}

\usepackage{multirow}
\usepackage{comment}
\usepackage{float}

\usepackage{amsmath,amssymb}

\usepackage{changepage}

\usepackage{textcomp,marvosym}

\usepackage{cite}

\usepackage{nameref,hyperref}

\usepackage[right]{lineno}

\usepackage[nopatch=eqnum]{microtype}
\DisableLigatures[f]{encoding = *, family = * }

\usepackage[table]{xcolor}

\usepackage{array}

\newcolumntype{+}{!{\vrule width 2pt}}

\newlength\savedwidth

\raggedright
\usepackage[aboveskip=1pt,labelfont=bf,labelsep=period,justification=raggedright,singlelinecheck=off]{caption}

\makeatletter
\renewcommand{\@biblabel}[1]{\quad#1.}
\makeatother

\usepackage{lastpage,fancyhdr,graphicx}
\usepackage{epstopdf}
\fancyheadoffset[L]{2.25in}
\fancyfootoffset[L]{2.25in}
\usepackage{xspace}

\usepackage[amssymb]{SIunits}
\addunit{\TCID}{TCID_{50}}
\addunit{\PFU}{PFU}
\addunit{\dpi}{dpi}
\addunit{\perh}{\hour^{-1}}

\begin{document}

\newcommand{\fixme}[1]{\textbf{\color{red}FIXME: #1}}
\newcommand{\ddt}[1]{\frac{\mathrm{d}#1}{\mathrm{d}t}}
\newcommand{\sinf}{\mathrm{inf}}
\newcommand{\RNA}{\mathrm{RNA}}
\newcommand{\Pri}{\mathcal{P}_\text{rior}}
\newcommand{\Po}{\mathcal{P}_\text{ost}}
\newcommand{\Lik}{\mathcal{L}}
\newcommand{\Dil}{\mathcal{D}\text{il}}
\newcommand{\pars}{\boldsymbol{\pi}}
\newcommand{\alphaB}{\alpha_{\frac{\text{BHK}}{\text{A549}}}}
\newcommand{\ssin}{\mathrm{SIN}}

\vspace*{0.2in}

\begin{flushleft}
{\Large
\textbf\newline{Mechanistic mathematical model of the in vitro infection dynamics of Bunyamwera and Batai viruses including MOI-dependent shortening of the eclipse phase}
}
\newline
\\

Bevelynn Whaler\textsuperscript{1 \Yinyang \ *},
Eleanor Todd\textsuperscript{2 \Yinyang},
Amelia B. Shaw\textsuperscript{2 \Yinyang},
John N. Barr\textsuperscript{2},
Catherine A. A. Beauchemin\textsuperscript{3, 4},
Grant Lythe\textsuperscript{1},
Carmen Molina-Par\'is\textsuperscript{1,5},
Mart\'in L\'opez-Garc\'ia\textsuperscript{1 *}

\bigskip
\textbf{1} Department of Applied Mathematics, School of Mathematics, University of Leeds, Leeds, United Kingdom
\\
\textbf{2} School of Molecular and Cellular Biology, Faculty of Biological Sciences, University of Leeds, Leeds, United Kingdom
\\
\textbf{3} Department of Physics, Toronto Metropolitan University, Toronto, Canada
\\
\textbf{4} RIKEN iTHEMS, Wako, Saitama, Japan
\\
\textbf{5} T-6, Theoretical Biology and Biophysics, Theoretical Division, Los Alamos National Laboratory, Los Alamos, NM, USA
\\
\bigskip

\Yinyang \ These authors contributed equally to this work.

* B.Williams1@leeds.ac.uk, M.LopezGarcia@leeds.ac.uk

\end{flushleft}

\section*{Abstract}
We develop a deterministic mathematical  model to quantify the distinct \textit{in vitro} infection dynamics of Bunyamwera virus (BUNV) and Batai virus (BATV) in A549 cells, incorporating cell division and natural death, continued entry of virions into already-infected cells, and shortening of the eclipse phase driven by re-infection. The model parameters were estimated making use of viral decay data, growth curves at two different inoculum concentrations, and extra-cellular genome copy measurements (for BUNV) via Markov chain Monte Carlo. Genome copy measurements were essential for constraining estimates of the number of cells that can become infected per unit of infectious virus for BUNV. We found that BUNV exhibited substantially longer eclipse and infectious periods than BATV, while BATV showed a higher per-cell virus production rate. Re-infection was predicted to shorten the eclipse phase for both viruses, but the effect was markedly stronger for BUNV. Together, these results provide a quantitative comparison of the \textit{in vitro} viral kinetics of BUNV and BATV and reveal substantial differences in their replication dynamics.

\section*{Author summary}
In this study, we combine laboratory experiments with mathematical modelling to quantify the \textit{in vitro} infection dynamics of two closely related vector-borne bunyaviruses, Bunyamwera virus (BUNV) and Batai virus (BATV).
By estimating model parameters making use of experimental data of infectious viral titres, cell numbers, and viral genome copies, we were able to constrain and compare the viral
life cycle properties of each of these viruses, such as how quickly cells become infected, how long cells take to begin producing virus, and for how long they continue releasing new viral particles. Our results show that BUNV and BATV kinetics differ substantially. We also found that infection of a cell by multiple viral particles at once may shorten the time it takes a cell to start producing viral particles.
This work combines mathematical modelling with experimental data to provide quantitative insight into processes that shape the replication dynamics of these viruses and which are difficult to measure directly.

\section*{Introduction}

Bunyamwera virus (BUNV) and Batai virus (BATV) are vector-borne orthobunyaviruses belonging to the \textit{Peribunyaviridae} family within the \textit{Bunyaviricetes} class. These are enveloped viruses that replicate in the cytosol and bud at Golgi membranes \cite{barker2023mechanisms}. The genome of each of these viruses consists of three negative-sense RNA segments, named small (S), medium (M), and large (L), reflecting their relative nucleotide 
length~\cite{walter2011recent}.
The S segment encodes the viral nucleoprotein (N) and a non-structural protein (NSs).
The M segment encodes a glycoprotein precursor, which is co-translationally cleaved by host-cell proteases into two surface glycoproteins (Gc and Gn), and it also encodes a non-structural protein (NSm).
The L segment encodes the RNA-dependent RNA polymerase (RdRp).

BUNV is considered the prototype member of the \textit{Peribunyaviridae} family and the \textit{Orthobunyavirus} genus, and has been widely used to study aspects of the molecular biology of these viruses, including 
replication~\cite{lowen2005attenuation, shi2010visualizing}, assembly~\cite{novoa2005key, shi2006requirement}, and transmission between vertebrate and arthropod hosts~\cite{terhzaz2025nsm}.
BATV, although closely related, exhibits slightly different phenotypic and epidemiological characteristics~\cite{zoller2024innate}.
These viruses have the potential to undergo reassortment when they co-infect the same host cell, a process during which gene segments are exchanged between viruses, forming new genotypes~\cite{lowen2018s}. This can lead to novel pathogens of public health and veterinary concern. 
For example, a natural reassortant between BATV and BUNV is the Ngari virus (NRIV), which carries the L and S segments of BUNV and the M segment of 
BATV~\cite{briese2006batai}. NRIV shows increased pathogenicity compared to its parental viruses and is associated with hemorrhagic fever in humans, although the molecular mechanism for this is still being studied~\cite{dutuze2021comparative, heitmann2021mammals, bowen2025probing}.
Therefore, quantifying the infection dynamics of BUNV and BATV is important to interpret their replication strategies and support future work to understand 
host-pathogen interactions and the potential emergence of pathogenic reassortants.

Mathematical modelling is a powerful tool for quantifying key processes of \textit{in vitro} viral infections, including cell infection rates, intra-cellular delays, viral production, and loss of virus infectivity.
It has been widely applied to RNA viruses such as influenza viruses~\cite{paradis2015impact, liao2016validating}, respiratory syncytial virus~\cite{beauchemin2019uncovering}, and Ebola virus~\cite{liao2020quantification}, 
since it allows to estimate parameters that are difficult or impossible to measure directly, to interpret viral growth curves, and to improve our understanding of how experimental conditions influence infection outcomes.
However, a mathematical model (MM) has not previously been used to study the \textit{in vitro} infection dynamics of orthobunyaviruses such as BUNV and BATV.

In this study, we conducted experiments to characterise the progression of BUNV and BATV infection in A549 cells and obtained time-course measurements of infectious viral titre and cell numbers to calibrate an MM.
For BUNV, we also obtained data on extra-cellular genome segment copy numbers.
We developed a mechanistic MM 
that extends existing viral dynamics MMs to capture features particularly relevant for these viruses and cells. In particular, we incorporated cell division and natural death, continued entry of virions into already-infected cells, and shortening of the eclipse phase driven by re-infection. 
Estimation of the MM's parameters is performed using Markov chain Monte Carlo (MCMC), allowing us to quantify uncertainty and explore parameter correlations.
Our results reveal pronounced differences between BUNV and BATV in eclipse-phase duration, infectious lifespan, and sensitivity to re-infection, providing new insight into their replication kinetics.

The objective of this work is to use a quantitative framework to 
provide estimates of mechanistic life cycle parameters and identify viral traits that differ substantially between BUNV and BATV. These insights can support the design of new experiments and guide hypotheses about viral replication strategies and reassortment dynamics of these segmented RNA viruses.

\section*{Results}

To quantify the \textit{in vitro} infection dynamics of BUNV and BATV, we infected A549 cells with BUNV or BATV at two different multiplicities of infection (MOI). In these experiments, we measured infectious viral titre and number of cells over time 
(see Fig~\ref{fig:posterior_predictions_infection}). In the BUNV higher MOI experiment, we also quantified the number of extra-cellular negative-sense RNA copies of each segment 
(see Fig~\ref{fig:posterior_predictions_RNA}). Since infectious virus will naturally decay outside of a cell in the supernatant and become non-infectious (this is described by the viral decay rate, $c$, in the MM), we also conducted viral decay experiments to help identify the decay rate for each virus 
(see Fig~\ref{fig:posterior_predictions_viral_decay}). The details of each experiment are provided in the Methods section.

\begin{figure}[!ht]
\begin{adjustwidth}{-2.25in}{0in}
\includegraphics[width = 7.5in]{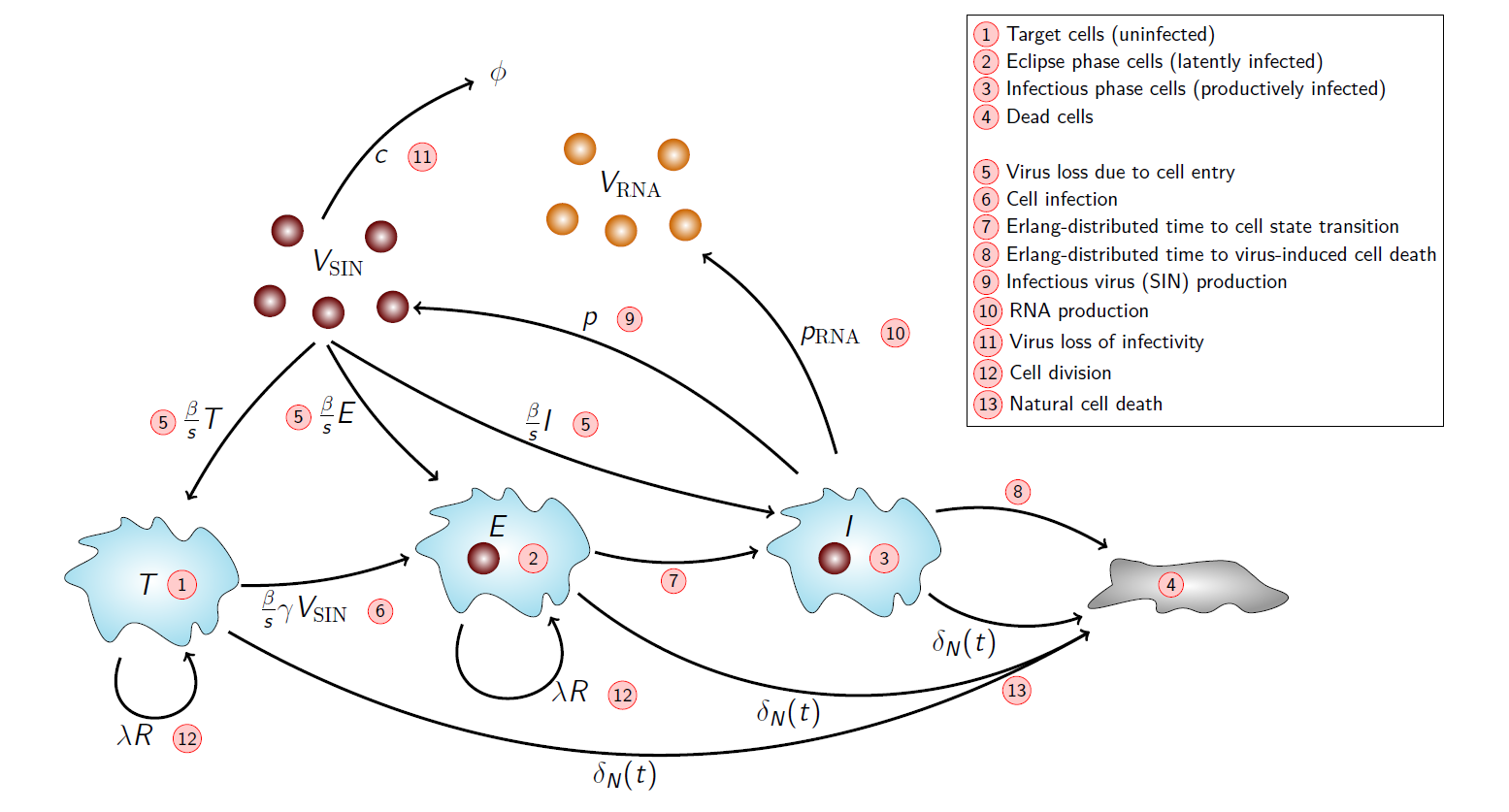}
\caption{{\bf Diagram of the \textit{in vitro} BUNV or BATV infection MM.} Uninfected cells, $T$, and eclipse phase cells, $E$, divide at rate $\lambda R$, where $R(t)$ represents the availability of resources for cell division in the media, which are assumed to be used up at a constant rate per cell. The time-dependent rate of natural cell death is given by $\delta_N(t)$, which is the hazard function of an Erlang-distributed random variable.
We consider entry of virions into all cells, $T$, $E$, and $I$. Each time an eclipse phase cell is re-infected by an additional virion, the Erlang distribution for the time to go from the eclipse phase to the infectious phase is shortened by jumping ahead to a future eclipse stage (see Fig~\ref{fig:model_diagram_jumping}).
See the Methods section for a complete description of the MM.}
\label{fig:model_diagram}
\end{adjustwidth}
\end{figure}

The MM used here to describe the \textit{in vitro} BUNV and BATV infection kinetics is shown 
in Fig~\ref{fig:model_diagram} and described in detail in the Methods section. In our \textit{in vitro} experiments, the amount of infectious virus in the supernatant was measured via an endpoint dilution (ED)
assay, which determines the fraction of wells that become infected (identified by the amount of cytopathic effect) when inoculated with a virus sample that has been diluted to different extents. In our MM, the unit that we use for infectious virus in the supernatant is the number of specific infections 
(SIN)~\cite{midSIN,quirouette2024does}. One SIN can be thought of as the dose of infectious virus that is needed to establish a detectable infection in an ED assay well, under the specific conditions of the ED assay used; that is, an observed infected well in the ED assay indicates that at least one SIN was inoculated into the well~\cite{midSIN,quirouette2024does}. However, one SIN may correspond to multiple infectious virions, so that the loss of one SIN due to cell entry may correspond to multiple cells becoming infected.
Therefore, we incorporate a parameter, $\gamma$, into the MM, which is defined as the number of cells that become infected per SIN lost to cell 
entry~\cite{quirouette2024does}. See 
Table~\ref{table:parameters} for a full list of the model parameters.

\begin{table}[!ht]
\begin{adjustwidth}{-2.25in}{0in}
\centering
\caption{
{\bf MM parameters~\eqref{eq:model_ODEs}, along with their descriptions, units, and prior distributions.}}
\begin{tabular}{|c|p{90mm}|c|c|}
\hline
\rule{0pt}{3ex}
Parameter & Description & Unit & Prior\\[0.3em]
         \hline
         \rule{0pt}{3ex}
         $V_0^\text{decay}$& SIN per well at time \unit{0}{h} in the BUNV viral decay experiment & SIN & $7.9\times 10^4$ (fixed)\\
         $V_0^\text{low}$& SIN per well at time \unit{0}{h} in the BUNV low MOI experiment & SIN & $9.7\times10^4$ (fixed)\\
         $V_0^\text{high}$& SIN per well at time \unit{0}{h} in the BUNV high MOI experiment & SIN & $4.5\times 10^6$ (fixed)\\[0.3em]
         $V_0^\RNA$& RNA per well at time \unit{0}{h} in the BUNV high MOI experiment & RNA & $3.6\times 10^7$ (fixed)\\[0.3em]
         \hline
         \rule{0pt}{3ex}
         $V_0^\text{decay}$& SIN per well at time \unit{0}{h} in the BATV viral decay experiment & SIN & $3.3\times 10^5$ (fixed)\\
         $V_0^\text{low}$& SIN per well at time \unit{0}{h} in the BATV low MOI experiment & SIN & $3.7\times10^5$ (fixed)\\
         $V_0^\text{high}$& SIN per well at time \unit{0}{h} in the BATV high MOI experiment & SIN & $2.7\times 10^7$ (fixed)\\[0.3em]
         \hline
         \rule{0pt}{3ex}
         $s$& volume of supernatant per well & \unit{}{\milli\litre} & $2$ (fixed)\\
         $\lambda$& initial cell division rate&h$^{-1}$& $0.048$ (fixed)\\
         $\tau_N$& average time to natural cell death &h& $174$ (fixed)\\
         $n_N$& number of natural cell death stages & - & $5$ (fixed)\\
         $\mu$& rate of resource consumption&ml$\cdot$(cell$\cdot$h)$^{-1}$& $4.8\times 10^{-8}$ (fixed)\\[0.3em]
         \hline
         \rule{0pt}{3ex}
         $T_0$& initial number of cells per well in the infection experiments & cell & log$_{10}T_0 \sim U(-\infty,\infty)$\\
         $\beta$& rate of viral cell entry &$\unit{}{\milli\litre}\cdot(\textrm{cell}\cdot \textrm{h})^{-1}$&log$_{10}\beta \sim U(-\infty, \infty)$\\
         $c$&rate of viral decay in supernatant&$\textrm{h}^{-1}$& log$_{10}c \sim U(-\infty, \infty)$\\
         $\gamma$& number of cells infected per SIN entry&$\textrm{cell}\cdot \textrm{SIN}^{-1}$&log$_{10}\gamma \sim U(-\infty,\infty)$\\
         $p$ & rate of SIN production by an infectious cell &$\ssin \cdot (\textrm{cell}\cdot\textrm{h})^{-1}$&log$_{10}p \sim U(-\infty, \infty)$\\
         $p_\RNA$ & rate of RNA production by an infectious cell &$\RNA \cdot (\textrm{cell}\cdot\textrm{h})^{-1}$&log$_{10}p_\RNA \sim U(-\infty, \infty)$\\
         $n_E$&number of eclipse phase stages&-& $n_E \sim U(0, 60)$\\
         $n_I$&number of infectious phase stages&-& $n_I \sim U(0, 60)$\\
         $\kappa$& number of eclipse stages jumped due to re-infection&-& $\kappa \sim U(0, 60)$\\
         $\tau_E$&average time of eclipse phase&h& $\tau_E \sim U(0, \infty)$\\
         $\tau_I$&average time of infectious phase&h& $\tau_I \sim U(0, 500)$\\
         $\omega$&fraction of virus in the supernatant remaining after washing&-&log$_{10}\omega\sim U(-\infty, 0)$\\
         $\alphaB$& number of SIN in BHKs per SIN in A549s&$\textrm{SIN}_\text{BHK}\cdot \textrm{SIN}^{-1}$& log$_{10}\alphaB \sim U(-\infty, \infty)$\\
\hline
\end{tabular}
\label{table:parameters}
\end{adjustwidth}
\end{table}

\begin{figure}[!ht]
\begin{adjustwidth}{-2.25in}{0in}
\includegraphics[width = 7.5in]{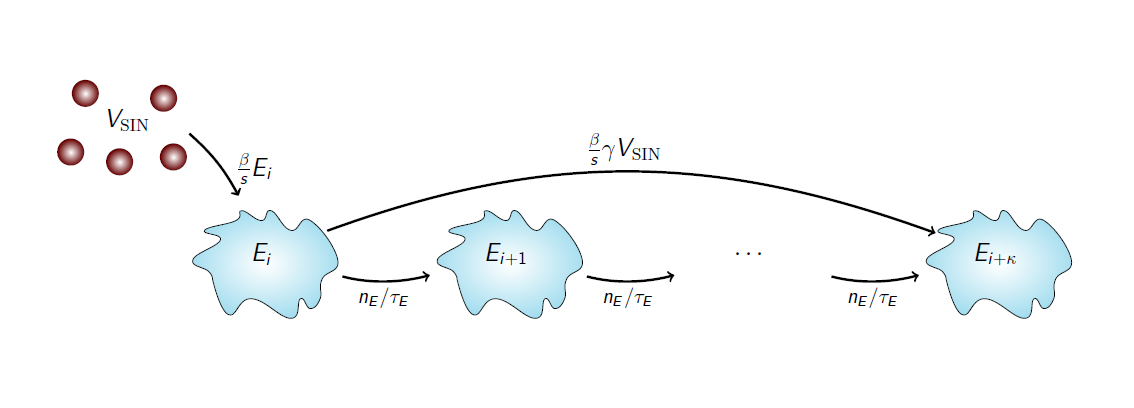}
\caption{{\bf Modelling a reduction in the length of the eclipse phase due to re-infection.} Each time an eclipse phase cell is re-infected by an additional virion, the Erlang distribution for the time to transition from the eclipse phase to the infectious phase is shortened by jumping ahead to a future eclipse stage. The parameter $\kappa$ defines the size of the jump taken ({\em i.e.}, how many eclipse stages are skipped each time a new virion enters).
}\label{fig:model_diagram_jumping}
\end{adjustwidth}
\end{figure}

There are three main differences in our MM compared with existing MMs of \textit{in vitro} viral dynamics, which allow it to capture key observed features of our experimental data: 
\begin{enumerate}

\item Motivated by the observed behaviour of our \textit{in vitro} A549 cell cultures, which exhibited an initial phase of cell population growth followed by decline, even in the mock experiment in which cells were not infected by virus (see Fig~\ref{fig:mock_cells_prediction}), we explicitly model cell division and natural death. To describe cell division, we define a variable that represents the resource availability  for cell division in the media, starting at $R(0)=1$ at the time when the well is inoculated with virus. Cells are assumed to use up resources at a constant rate per cell, and the cell division rate is proportional to $R(t)$. Therefore, the division rate decreases over time as resources are used up. To describe natural cell death, we use a time-dependent rate defined relative to the start of the experiment. This choice reflects the assumption that population-level factors, such as resource depletion and waste accumulation, influence cell survival over time.
We estimate the model parameters that describe the processes of resource-limited cell division and 
time-dependent natural death using data from the mock infection experiments (see the Methods section).

\item 
While some viruses, such as influenza A, exhibit superinfection exclusion, this has not yet been investigated for BUNV or BATV, and experimental studies indicate that cellular co-infection of bunyaviruses is possible~\cite{bermudez2022incomplete}. Therefore, in the absence of detailed data or a hypothesised mechanism of superinfection exclusion for BUNV and BATV in this experimental setting, we make the assumption that
virus can continue to enter cells throughout the course of infection, rather than
assuming a finite window for virus entry into cells which closes after initial infection.
We implement this in the MM by adapting the removal rate of extra-cellular virus due to cell entry.

\item
In our BUNV experiments, we observed that virus started to be produced noticeably earlier in the higher MOI experiment than the lower MOI experiment (see Fig~\ref{fig:posterior_predictions_infection}, panels C1 and C2), which could not be captured by a standard \textit{in vitro} viral dynamics MM. We hypothesise that if a cell becomes infected by multiple virions, the increase in intra-cellular viral load (more templates for genome replication and protein synthesis) could increase the initial rate of virus replication within the cell. Therefore, additional virion entry events after a cell has become infected (entered the eclipse phase, $E$ in 
Fig~\ref{fig:model_diagram})
may lead to a reduction in the time it takes a cell
to release progeny virus (enters the infectious phase, $I$ in Fig~\ref{fig:model_diagram}). Note that the period after virus entry but before the release of virions is called the eclipse phase, and the period during which a cell is releasing progeny virions is called the infectious phase. The MM incorporates a multi-stage representation of
the eclipse phase, so that the time a cell takes to start releasing virus follows an Erlang distribution. We implement the possible shortening of the eclipse phase due to re-infection by assuming that
each time an additional virion infects a cell, the time that the cell spends in the eclipse phase is shortened by jumping ahead to a future eclipse stage (see Fig~\ref{fig:model_diagram_jumping}).
The parameter $\kappa$ defines the size of the jump taken ({\em i.e.}, how many eclipse stages are skipped each time a new virion enters).
\end{enumerate}

For each virus separately, the parameters of the MM in Fig~\ref{fig:model_diagram} were estimated by simultaneously using the viral titre measurements and cell counts from the infection experiments
(Fig~\ref{fig:posterior_predictions_infection}), the number of extra-cellular copies of each genome segment (BUNV higher MOI experiment only, 
Fig~\ref{fig:posterior_predictions_RNA}), and the viral titre measurements from the viral decay experiments
(Fig~\ref{fig:posterior_predictions_viral_decay}). We used Markov chain Monte Carlo (MCMC) to estimate the posterior distribution of the parameter vector, $\pars = (T_0, \beta, c, \gamma, p, p_\RNA, n_E, n_I, \kappa, \tau_E, \tau_I, \omega, \alphaB)$. Other parameters and initial conditions were fixed to the values provided in Table~\ref{table:parameters}.
Details about the MCMC algorithm used are provided in the Methods section.

Fig~\ref{fig:posterior_predictions_infection} shows (for the lower and higher MOI experiments for each virus) the posterior MM predictions of the number of total cells, uninfected cells, eclipse phase cells, infectious cells, and the amount of extra-cellular SIN, along with the experimental observations used in the parameter inference (dots showing the number of cells and most likely number of SIN per well, given the ED assay outcomes, for each time point and replicate). The parameter set corresponding to the largest posterior probability has been used to obtain the lines. The shaded regions show the 95\% credible intervals (CI) of the MM predictions, based on $10^3$ parameter sets sampled at random with replacement from the $26 \times 10^3$ accepted parameter sets.

\begin{figure}[!ht]
\begin{adjustwidth}{-2.25in}{0in}
\centering
\includegraphics[width = 7.1in]{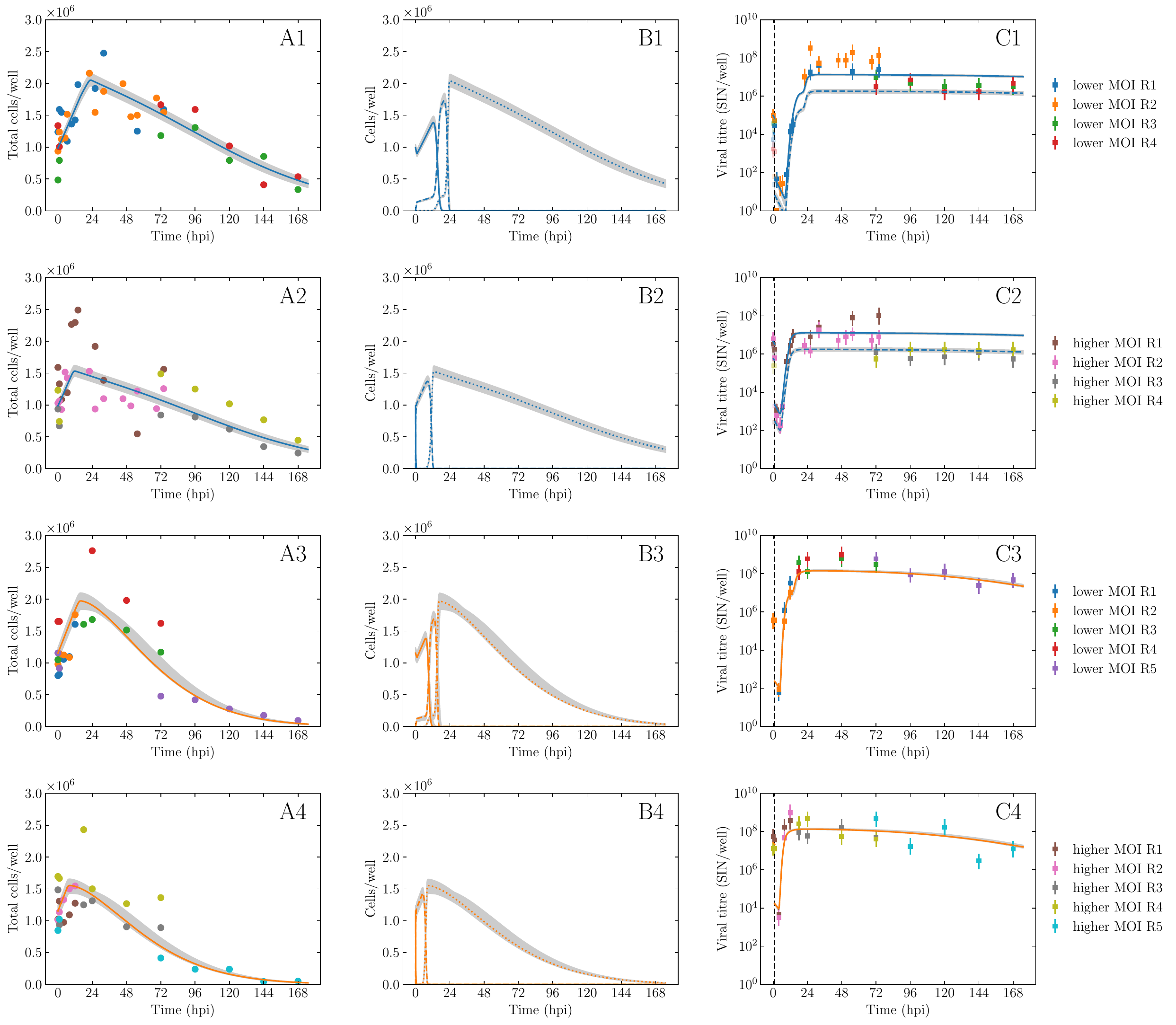}
\caption{{\bf Posterior MM predictions compared with the cell number and viral titre data used in the MCMC for parameter inference.}
Lines show predictions corresponding to the most likely parameter set and shaded regions show the 95\% credible intervals (CI) of the predictions based on $10^3$ parameter sets sampled at random with replacement from the $26 \times 10^3$ accepted parameter sets. (A1-C1) BUNV lower MOI experiment;
(A1) Predicted total remaining cells ($T+E+I$) compared with experimentally measured number of cells per well (counted using a hemocytometer after collecting the supernatant, washing once with 1$\times$ PBS, and detaching remaining cells with trypsin); (B1) Predicted number of cells in the uninfected (T, solid line), eclipse (E, dashed line), and infectious (I, dotted line) states; (C1) Predicted number of SIN/well. The solid line corresponds to SIN measured using A549 cells in the ED assay (R1 and R2) and the dashed line corresponds to SIN measured using BHK cells in the ED assay (R3 and R4). The viral titre data points and error bars correspond to the most likely SIN per well and 95\% CI estimated by 
midSIN~\cite{midSIN}, but the likelihood used in the MCMC was based on the full ED assay outcome via 
Eq~\eqref{eqn:lik_ED}~\cite{quirouette2024does}; (A2-C2) Same as (A1-C1) but for the BUNV higher MOI experiment; (A3-C3) Same as (A1-C1) but for the BATV lower MOI experiment and all viral titres were measured using A549 cells; (A4-C4) Same as (A3-C3) but for the BATV higher MOI experiment.}
\label{fig:posterior_predictions_infection}
\end{adjustwidth}
\end{figure}

\begin{figure}[!ht]
\centering
\includegraphics[width = 0.9\textwidth]{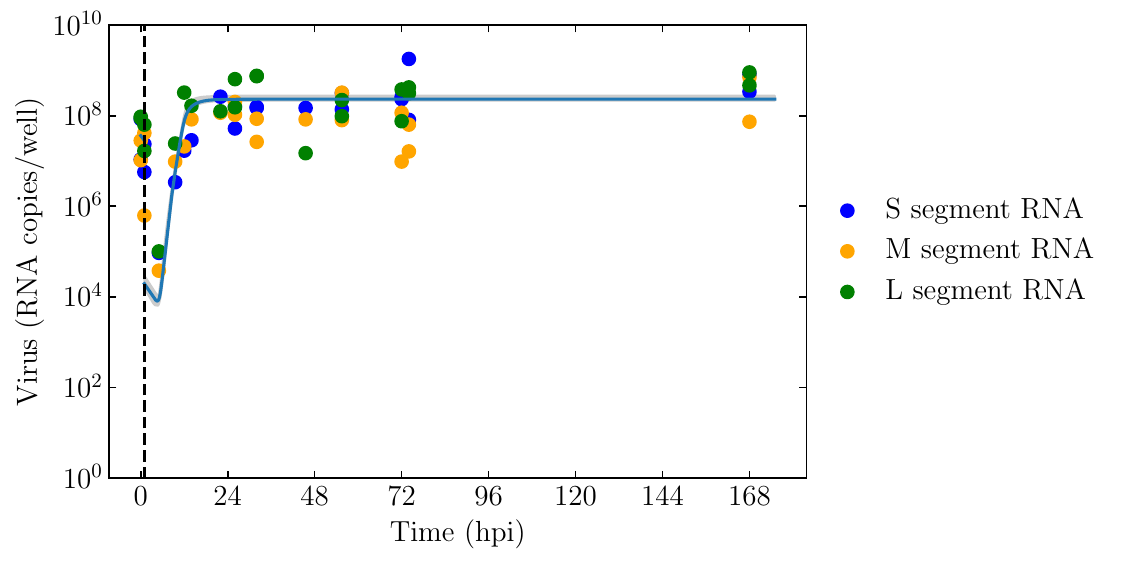}
\caption{{\bf Posterior MM predictions for the number of RNA copies of each segment, $V_\RNA(t)$, in the high-MOI BUNV experiment, compared with the observed genome counts which were used in the parameter inference.}}
\label{fig:posterior_predictions_RNA}
\end{figure}

A previous modelling study using similar experimental data on \textit{in vitro} infection of A549 cells with a seasonal
influenza A virus strain found that the parameters representing the number of cells that become infected per SIN lost to cell entry ($\gamma$) and the rate at which SIN enters cells ($\beta$) were poorly constrained, and a lower bound on $\beta$ had to be imposed~\cite{quirouette2024does}. 
For BUNV, we also found that without incorporating the extra-cellular genome copy measurements into the parameter inference, $\gamma$ and $\beta$ remained practically unidentifiable. That is, the viral titre measurements alone could be matched with arbitrarily large $\gamma$ and correspondingly small $\beta$, since it is the product $\gamma\beta$ that determines the effective rate at which cells become infected in the MM~\cite{quirouette2024does}.
However, including the genome-copy measurements from the high-MOI BUNV experiment imposes an additional biological constraint on the total number of virions present in the supernatant, and therefore provided an upper bound on the number of cells that become infected per SIN lost to cell entry 
(represented by $\gamma$). We included the requirement that the maximum number of cells that could become infected by the infectious virus in the supernatant ($\gamma V_\ssin$) must be smaller than the number of RNA copies of a given segment ($V_\RNA$). This is based on the assumption that the number of cells that can become infected is limited by the number of infectious virions in the supernatant, and each infectious virion must contain at least one copy of each segment.
This therefore enabled us to reject the high-$\gamma$/low-$\beta$ parameter combinations and constrain the posterior distributions for $\gamma$ and $\beta$ to biologically plausible ranges, highlighting the value of this type of data when calibrating \textit{in vitro} viral dynamics MMs.
Fig~\ref{fig:posterior_predictions_RNA} compares the posterior MM predictions of the number of RNA copies in the supernatant ($V_\RNA$) with the measurements of extra-cellular viral genome copies for each segment in the high-MOI BUNV experiment. The agreement between the MM predictions and the observed genome copy trajectories indicates that the inferred parameters are consistent with realistic levels of total virus production, as well as reproducing the observed infectious viral titre dynamics in 
Fig~\ref{fig:posterior_predictions_infection}. Importantly, reproducing viral titre dynamics alone does not guarantee realistic predictions of total virus abundance. Parameter sets with a large value of $\gamma$, which would be accepted into the posterior sample if we did not include the genome copy data, would correspond to a much larger  total virion
number, and therefore would not provide a realistic prediction of the number of genome copies.

Fig~\ref{fig:posterior_predictions_viral_decay} compares the posterior MM predictions of the viral decay curves (in the absence of cells) with the viral titre data sets from the viral decay experiments which were used in the parameter inference. The solid lines show the prediction corresponding to the most likely value (which maximises the posterior density) of the viral infectivity decay rate ($c$) in the posterior sample for each virus, and the shaded regions show the 95\% CI of the MM predictions.

Fig~\ref{fig:posteriors} compares the marginal posterior distributions of the parameters that were estimated for each virus, while the medians and 95\% credible intervals are reported in Table~\ref{table:params_comparison}. We found that it was possible to substantially constrain most of the model parameters, yet
a few parameters are estimated to be rather different between the two viruses.
One of the most notable differences is in the average eclipse period for singly infected cells ($\tau_E$). This is estimated to be around twice as long for BUNV than BATV (median of around \unit{15}{h} for BUNV compared to \unit{8}{h} for BATV). For both viruses, large values up to the prior bound of $n_E=60$ are accepted for the shape parameter of the Erlang-distributed eclipse period, predicting little variance in the time at which cells start to produce virus. 
We note that the results also predict that cells infected by more than one virion will have shorter eclipse phases, with each virion that enters an eclipse phase cell causing the total eclipse period to be reduced by around \unit{0.53}{h} and \unit{0.14}{h}, for BUNV and BATV, respectively (see left of 
Fig~\ref{fig:derived_posteriors}). This implies that the eclipse period is MOI-dependent, as seen in 
Fig~\ref{fig:comparison_across_MOIs}, where for BUNV, the time at which cells first become infectious (panel C) and start producing virus (panel D) is much earlier for high MOI (blue dashed line) than low MOI (blue solid line).

\begin{figure}[!htp]
\begin{adjustwidth}{-2.25in}{0in}
\centering
\includegraphics[width = 7in]{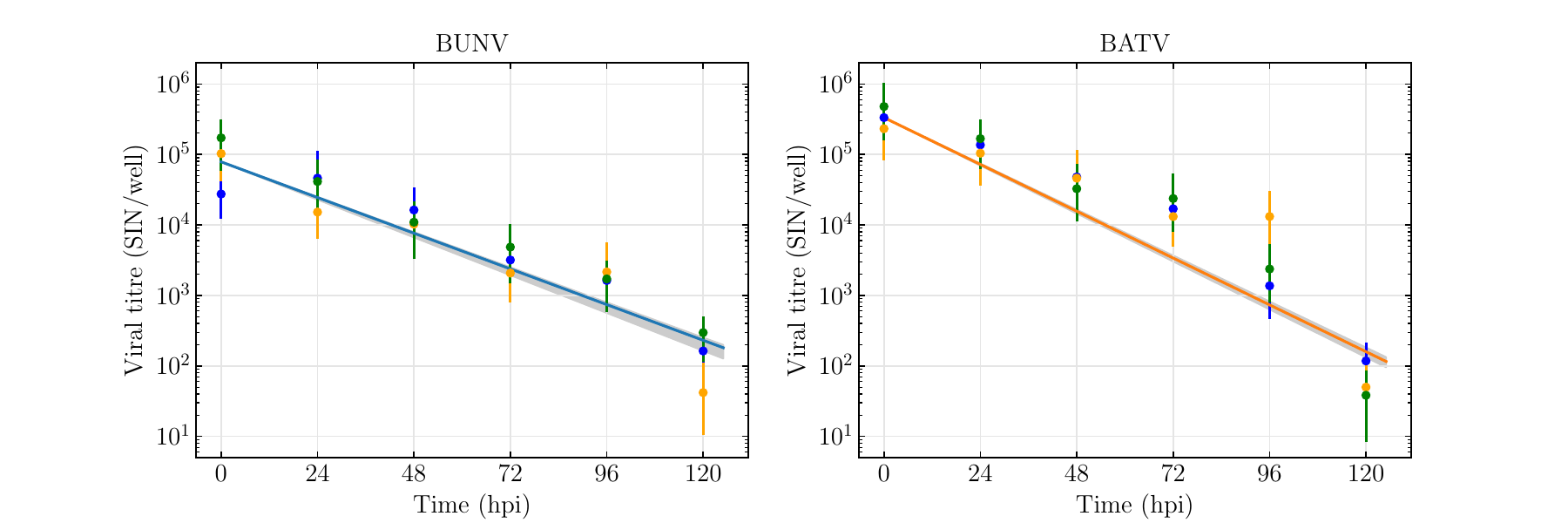}
\caption{{\bf Posterior MM predictions for the BUNV (left) and BATV (right) viral decay experiments, compared with the viral titre data from three replicates.}}
\label{fig:posterior_predictions_viral_decay}
\end{adjustwidth}
\end{figure}

Similarly to the eclipse period, the average infectious period ($\tau_I$) is estimated to be much longer for BUNV than BATV.
For BUNV, the median of the marginal posterior distribution indicates that infectious cells die after a mean time of around \unit{195}{h} (if they have not already died due to natural cell death) and the infectious period is predicted to be exponentially distributed ($n_I=1$). For BATV, the median estimate of the average time to virus-induced death is around $\tau_I = \unit{66}{h}$. However, the BATV posterior distribution for $\tau_I$ is bimodal because the MM can describe the data well with $n_I=2$ and $\tau_I \approx \unit{65}{h}$, or with $n_I=3$ and $\tau_I\approx \unit{70}{h}$.

The rate at which a viral particle enters
a cell ($\beta$) is estimated to be slightly higher for BUNV than BATV. Similarly, 
the number of cells that become infected per SIN lost to cell entry ($\gamma$) is slightly higher for BUNV than BATV; for BATV, each SIN that enters cells is estimated to result in $2.5$ infected cells (median estimate), whereas for BUNV, each SIN that enters cells is estimated to result in $5.8$ infected cells.
On the other hand, the production rate of virus by infectious cells ($p$) is lower for BUNV than BATV (around \unit{4}{SIN/h} for BUNV compared with around \unit{24}{SIN/h} for BATV).
However, since the duration of virus production ($\tau_I$) is predicted to be shorter for BATV, the estimated basic reproduction number, given by $\tau_I \, p \, \gamma \, \frac{\beta \, T_0}{\beta \, T_0 + c \, s}$, is slightly lower for this virus (see right of 
Fig~\ref{fig:derived_posteriors}), with median estimates of 3,960 for BUNV and 2,813 for BATV. We note that this expression for the basic reproduction number does not consider natural death of cells, which decreases the overall average duration of virus production from both BUNV- and BATV-infected cells.

In the posterior distributions for both viruses, there is a positive correlation between the rate at which infectious cells produce virus ($p$) and the rate at which the virus enters cells ($\beta$), which is shown in Fig~\ref{fig:p_vs_beta}. This arises from the MM assumption that viruses can continue to enter cells throughout the course of infection, rather than assuming a finite window for viral entry that closes after an initial infection.
As a consequence of this assumption, increasing the value of the viral entry rate ($\beta$) leads to a reduced viral titre plateau at later times, as more virus is removed from the supernatant through cell entry. However, this can be compensated for by a slightly higher viral production rate ($p$) in order for the MM prediction to capture the observed data.

\begin{figure}[htp]
\begin{adjustwidth}{-2.25in}{0in}
\includegraphics[width = 6.9in]{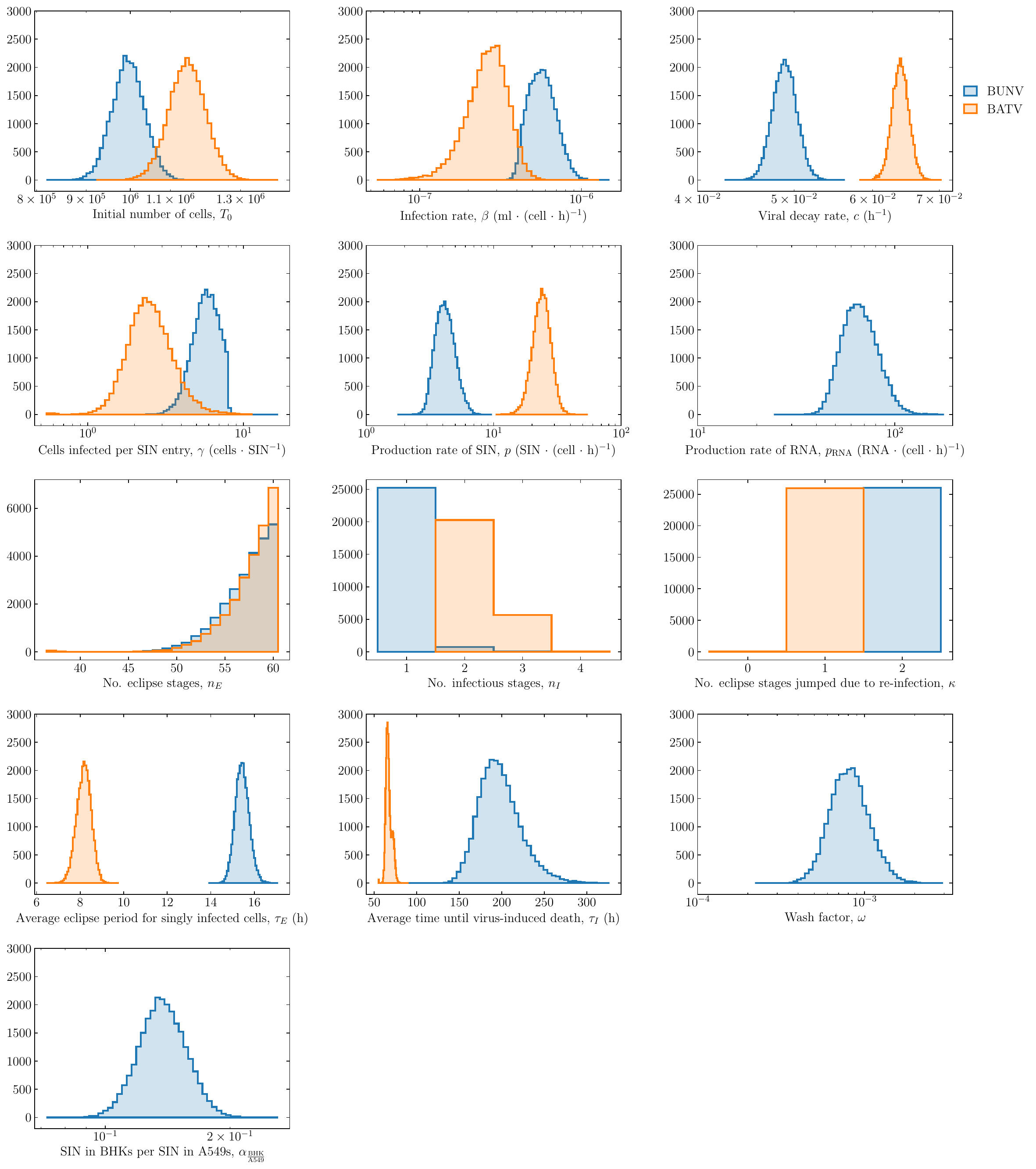}
\caption{{\bf Marginal posterior distributions of model parameters for BUNV (blue) and BATV (orange).} For each virus, the corresponding posterior distribution of parameters in the MM given in 
Eq~\eqref{eq:model_ODEs} was estimated based on viral titre data and cell numbers from the infection experiments with two different MOIs 
(Fig~\ref{fig:posterior_predictions_infection}), as well as viral titre data from the viral decay experiments (Fig~\ref{fig:posterior_predictions_viral_decay}). For BUNV, we also used measurements of the number of extra-cellular copies of each genome segment from the higher MOI experiment 
(Fig~\ref{fig:posterior_predictions_RNA}).}
\label{fig:posteriors}
\end{adjustwidth}
\end{figure}

\begin{table}[!ht]
    \begin{adjustwidth}{-2.25in}{0in}
    \centering
    \caption{{\bf Medians and 95\% credible intervals of the marginal posterior distributions for each parameter and each virus.}} 
    \begin{tabular}{|c|c|c|c|}
         \hline
         Parameter & Unit & BUNV & BATV\\[0.3cm]
         \hline
         \rule{0pt}{3ex}
         $T_0$ & cell 
         & $9.9\times 10^5$ ($9.2\times 10^5$, $1.1\times 10^6$)
         & $1.1\times 10^6$ ($1.0\times 10^6$, $1.2\times 10^6$) \\[0.3cm]
         
         $\beta$ & $\unit{}{\milli\litre}\cdot(\textrm{cell}\cdot \textrm{h})^{-1}$
         & $5.7\times 10^{-7}$ ($4.1\times 10^{-7}$ , $8.5\times 10^{-7}$)
         & $2.7\times 10^{-7}$ ($1.3\times 10^{-7}$ , $4.3\times 10^{-7}$) \\[0.3cm]

         $c$ 
         & $\textrm{h}^{-1}$ 
         & $4.9\times 10^{-2}$ ($4.6\times 10^{-2}$ , $5.2\times 10^{-2}$)
         & $6.4\times 10^{-2}$ ($6.2\times 10^{-2}$ , $6.7\times 10^{-2}$) \\[0.3cm]
         
         $\gamma$ 
         & $\textrm{cell}\cdot \textrm{SIN}^{-1}$
         & $5.8 \ (3.8, \ 7.8)$ 
         & $2.5 \ (1.4, \ 5.1)$ \\[0.3cm]
         
         $p$ 
         & $\textrm{SIN} \cdot (\textrm{cell}\cdot\textrm{h})^{-1}$
         & $4.1 \ (3.1, \ 6.0)$
         & $23.9 \ (17.1 \ 32.4)$ \\[0.3cm]

         $p_\RNA$ 
         & $\RNA \cdot (\textrm{cell}\cdot\textrm{h})^{-1}$
         & $65.8 \ (46.1, \ 100.8)$
         & N/A \\[0.3cm]

         $n_E$ 
         & -
         & $58 \ (51, \ 60)$
         & $58 \ (51, \ 60)$ \\[0.3cm]
         
         $n_I$ 
         & -
         & $1 \ (1, \ 2)$ 
         & $2 \  (2, \ 3)$ \\[0.3cm]
         
         $\kappa$ 
         & -
         & $2 \ (2, \ 2)$
         & $1 \ (1, \ 1)$ \\[0.3cm]
         
         $\tau_E$ 
         & h
         & $15.4 \ (14.8, \ 16.1)$ 
         & $8.2 \ (7.5, \ 8.8)$ \\[0.3cm]
         
         $\tau_I$ 
         & h
         & $195.4 \ (156.0, \ 255.1)$
         & $66.4 \ (62.0, \ 74.7)$ \\[0.3cm]
         
         $\omega$
         & -
         & $8\times 10^{-4} \ (5\times 10^{-4}, \ 10^{-3})$
         & $8\times 10^{-4}$ (fixed) \\[0.3cm]

         $\alphaB$ 
         & $\textrm{SIN}_\text{BHK}\cdot \textrm{SIN}^{-1}$
         & $1.4\times 10^{-1} \ (1.1\times 10^{-1}, \ 1.8\times 10^{-1})$
         & N/A \\[0.3cm]
         
         \hline
         
    \end{tabular}
    \label{table:params_comparison}
\end{adjustwidth}
\end{table}

\begin{figure}[!h]
\begin{adjustwidth}{-2.25in}{0in}
\vspace{5mm}
\includegraphics[width = 7.5in]{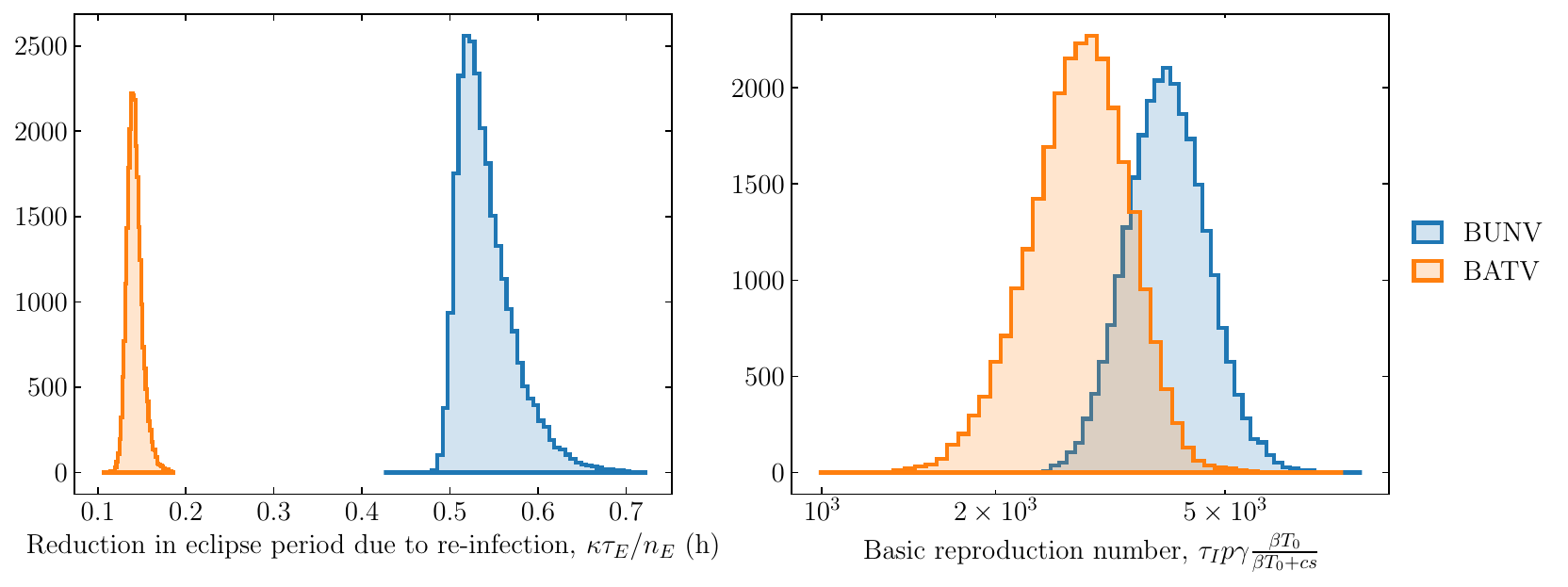}
\caption{{\bf Posterior distributions of relevant parameter combinations for each virus.} 
These are obtained by transformations of the posterior distributions in Fig~\ref{fig:posteriors}. Left: Distributions of the amount by which the average eclipse period is reduced for every additional virion that enters a cell.
Right: Distributions of the basic reproduction number (note that $s$ here is the volume of supernatant in one well, see Table~\ref{table:parameters}).}
\label{fig:derived_posteriors}
\end{adjustwidth}
\end{figure}

\begin{figure}[!h]
    \centering
    \includegraphics[width=0.7\textwidth]{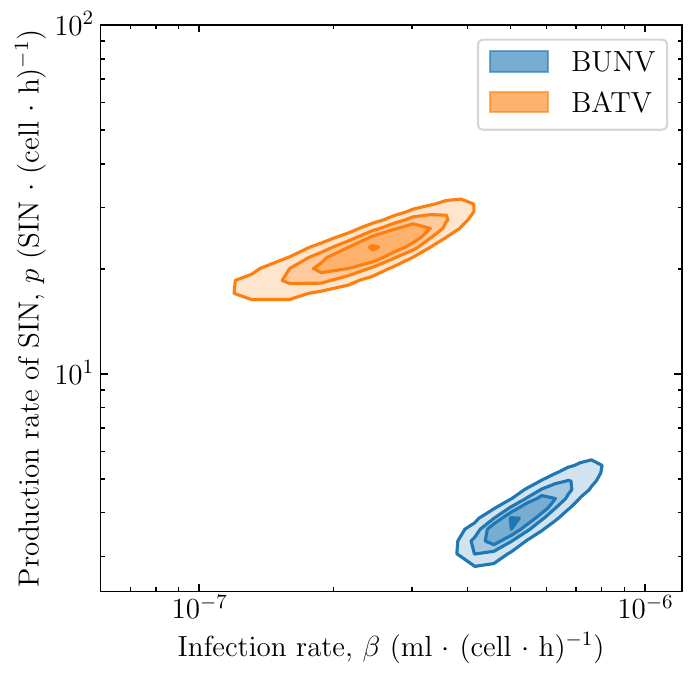}
    \caption{\textbf{
    Two-dimensional histograms showing  the relationship between the rate at which infectious cells produce virus ($p$) and the rate at which viruses enter cells ($\beta$) in the posterior sample for each virus.}}
    \label{fig:p_vs_beta}
\end{figure}

We have compared the posterior MM predictions for BUNV and BATV in Fig~\ref{fig:comparison_across_MOIs}, where two different MOIs are considered, assuming that there are initially $T_0=10^6$ cells per well in each case. For each virus, we use the posterior parameter distribution to simulate the scenario in which
(some amount of) infectious virus (in units of SIN) is added to the well; $V_\sinf(0)=10^5$ SIN and $V_\sinf(0)=7\times 10^6$ SIN for the low MOI (solid lines) and high MOI (dashed lines), respectively. At low MOI, BATV exhibits an earlier and steeper initial rise in viral titre due to a shorter eclipse phase and higher virus production rate per cell, whereas BUNV shows delayed virus production and a lower peak viral titre. Increasing the MOI leads to an earlier onset of virus production for both viruses, but with a much stronger effect observed for BUNV, consistent with the eclipse phase length having a greater sensitivity to re-infection. That is, BUNV has a larger inferred value of $\kappa$, representing the number of eclipse stages that are skipped each time an additional virion enters a cell, and each stage of the Erlang-distributed eclipse period corresponds to a larger period of time since the total average eclipse period ($\tau_E$) is greater for BUNV, but the number of eclipse stages ($n_E$) is similar for both viruses. At later time points, there is a clear decay in viral titre for BATV, whereas BUNV persists longer due to a more prolonged infectious period (greater number of infectious cells remaining at late times).

\begin{figure}[!ht]
\begin{adjustwidth}{-2.25in}{0in}
\centering
    \includegraphics[width=7.5in]{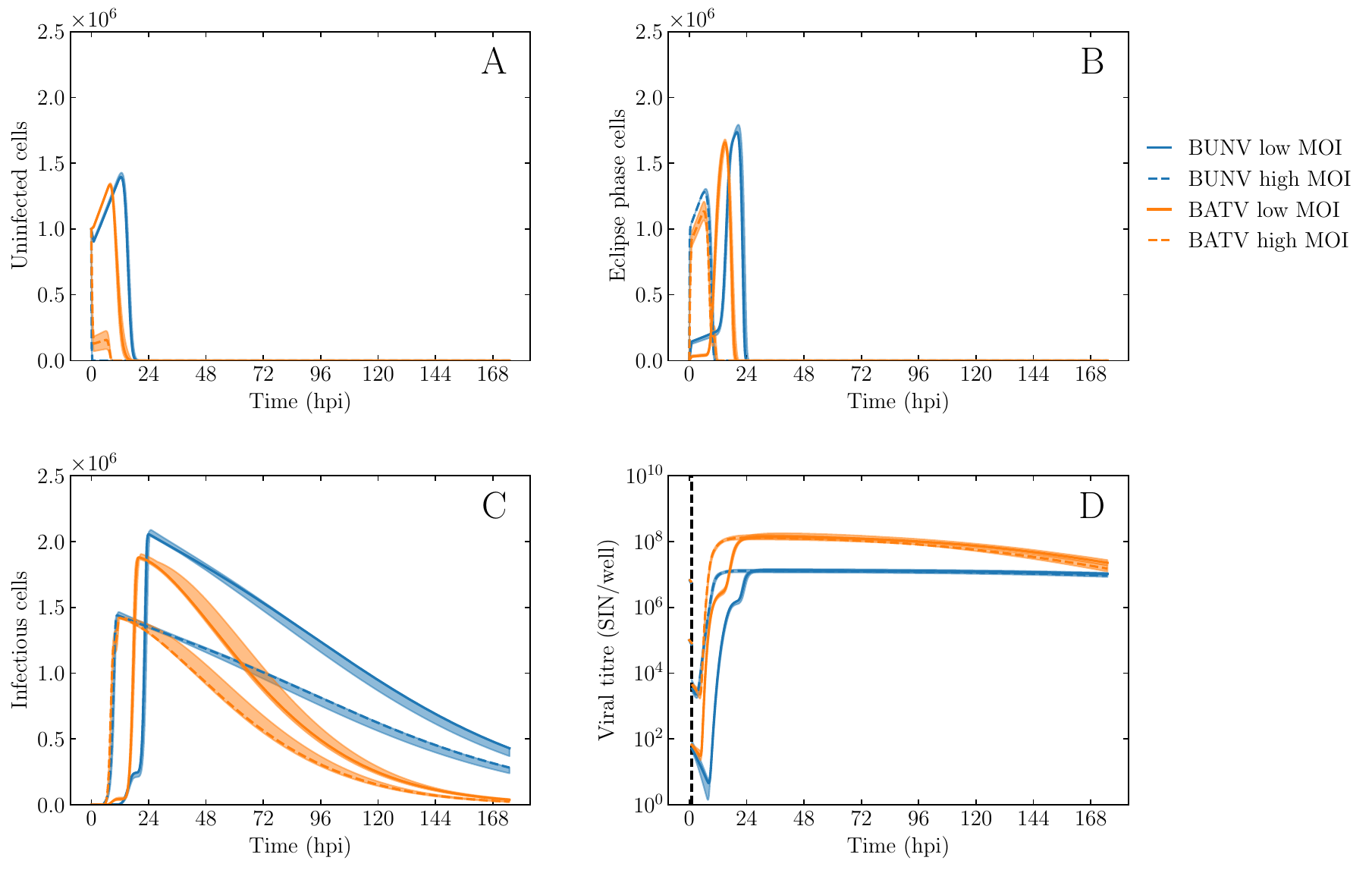}
    \caption{\textbf{Comparison of the predicted \textit{in vitro} infection dynamics of BUNV (blue) and BATV (orange) across MOIs.} Initial conditions used were $T(0)=10^6$ cells/well and either a low MOI of $V_\ssin(0)= 10^5$ SIN/well for each virus (solid lines) or a high MOI of $V_\ssin(0)= 7\times 10^6$ SIN/well for each virus (dashed lines). (A) Predicted number of cells in the uninfected state; (B) Predicted number of cells in the eclipse state; (C) Predicted number of cells in the infectious state; (D) Predicted viral titre in units of SIN/well.}
    \label{fig:comparison_across_MOIs}
    \end{adjustwidth}
\end{figure}

\section*{Discussion}

In this study, we conducted experiments to quantify the \textit{in vitro} infection dynamics of BUNV and BATV in A549 cells.
We used the resulting experimental measurements of infectious viral titre, cell numbers, and extra-cellular genome copies over time to develop and calibrate a mechanistic ordinary differential equation (ODE) model of the \textit{in vitro} infection dynamics of these two viruses.
In order to explain features of the experimental data that could not be captured by standard MMs of \textit{in vitro} viral dynamics, we extended the MM to incorporate resource-limited cell division and time-dependent natural death, continued virion entry into infected cells, and faster eclipse-phase progression driven by re-infection.
Using MCMC, we were able to estimate key infection parameters for both viruses and quantify differences in how the infections progress. This represents, to our knowledge, the first development and application of mathematical modelling to characterise the \textit{in vitro} dynamics of bunyaviruses, a group of segmented negative-sense RNA viruses of increasing public health and epidemiological importance~\cite{soldan2005emerging,hawman2025tick,tilston2024oropouche}.

Our results suggest that BUNV and BATV differ considerably in the length of the eclipse and infectious phases of their life cycles. BUNV-infected cells exhibited an eclipse period that was around twice as long as that of BATV, as well as a substantially longer infectious period. This difference aligns with previous experimental observations suggesting that BATV induces a stronger intra-cellular innate immune response than BUNV. It is thought that the NSm of BATV may trigger apoptosis, in contrast to the possible anti-apoptotic functions of the NSm and NSs of 
BUNV~\cite{zoller2024innate,kohl2003bunyamwera,leonard2006interaction}.

Another main finding of this study is that re-infection may accelerate intra-cellular infection progression by shortening the eclipse phase. The data from our experiments suggested that the duration before viral particles start to be released by cells is shorter at high MOI, but this could not be explained by standard viral dynamics MMs. Therefore, we included a mechanism in the MM to capture this effect and inferred how much the eclipse phase may be shortened by additional virions infecting an eclipse phase cell. Parameter inference predicted that an effect may be present for both viruses, but that the effect was stronger for BUNV than BATV.
Along with the longer delay before progeny virus is observed to be released, this could suggest that BUNV replicates more slowly than BATV within these cells, so that an increase in cellular MOI due to additional virion entry during the early stages of cell infection has a larger impact on the eclipse period. In future, measurements of the intra-cellular abundance of genome segments and viral proteins for both viruses could enable a comparison of the intra-cellular infection dynamics. The dynamics of co-infection between different orthobunyaviruses is also a key question given its role in reassortment and the potential for emergence of novel pathogenic strains~\cite{heitmann2021mammals}.
It is possible that co-infection could also alter the timescales of intra-cellular infection, which may influence the likelihood of reassortment events \textit{in vitro}.

Viral titre measurements alone are insufficient to constrain estimates of the number of cells that could become infected from one SIN (represented by $\gamma$). Indeed,  viral titre measurements could be matched with an arbitrarily large value of $\gamma$ and a correspondingly small infection rate ($\beta$), since it is the product $\gamma\beta$ that determines the overall rate at which cells become infected. This identifiability issue was identified in a previous \textit{in vitro} viral dynamics modelling study~\cite{quirouette2024does}, but it is only relevant when the MM includes a term representing the loss of extra-cellular virus due to cell entry (otherwise, a single parameter is typically used to represent the product $\gamma\beta$). Here, we were able to resolve this identifiability issue and obtain biologically plausible parameter estimates. We were able to do so by incorporating data from the high-MOI BUNV experiment on the number of extra-cellular copies of the L, M, and S genome segments, and placing a constraint on $\gamma$ by specifying that the maximum number of cells that could become infected by the infectious virus in the supernatant ($\gamma V_\ssin$) must be smaller than the number of RNA copies of a given segment ($V_\RNA$).

Our analysis provides new quantitative insight into the \textit{in vitro} replication dynamics of BUNV and BATV, but there are important uncertainties and limitations.
The MM used here assumes that infection spreads exclusively through released extra-cellular virus. This is a simplification, since the virus may also spread directly between cells. For lymphocytic choriomeningitis
arenavirus (LCMV), which belongs to the \textit{Arenaviridae} family (a different family within the \textit{Bunyaviricetes} class), observations have been made that are consistent with direct transfer of assembled virions or ribonucleoprotein genome segments between cells~\cite{byford2024lymphocytic}. Cell-to-cell viral transmission has also been reported for other RNA viruses that bud intra-cellularly, such as 
hepatitis C virus~\cite{timpe2008hepatitis,brimacombe2011neutralizing}. If such a mechanism occurs during infection with BUNV or BATV, it would increase the rate of cell infection independently of the extra-cellular viral titre, potentially altering several of the inferred parameter values.
However, incorporating cell-to-cell spread in the MM would require additional data capable of distinguishing between transmission mediated by free virus or cell-to-cell contact. For LCMV, blocking the extra-cellular route of infection by adding neutralising antibody to supernatants during infection revealed that around 50\% of transmission was via inter-cellular 
connections~\cite{byford2024lymphocytic}, but this has not yet been quantified for orthobunyaviruses.
Future work could focus on collecting additional data  sets to help discriminate between alternative MM structures ({\em e.g.}, with or without direct cell-to-cell viral transmission) that produce similar viral titre predictions, but differ substantially in their predictions of currently unobserved quantities such as the number of infected cells. Spatial data or measurements of intra-cellular markers such as nucleoprotein levels could be used to infer the fraction of infected cells, and it may be found that these data are not compatible with an MM in which infection spreads exclusively via extra-cellular virus.

As well as providing insight into the importance of direct cell-to-cell viral spread, information on the number of infected cells would be valuable to address another limitation of our approach. In the experiments, we inferred the total number of cells over time from counts of those remaining in the well after removal of the supernatant and washing, and we assumed that this corresponds directly to the total population of uninfected, eclipse-phase, and infectious cells ($T$+$E$+$I$). However, this is a simplification, since cells that have ceased virus production ({\em i.e.}, have left the infectious compartment, $I$) may remain adherent and be counted in the experimental cell measurements despite no longer contributing to viral output. This could lead to discrepancies between the MM predictions and the true distributions of cells across infection states. However, it was necessary for us to make this simplifying assumption since we found that the period of viral production from cells ($\tau_I$) was not identifiable from the viral titre data alone.

Another limitation is that the MM was not able to capture the peak viral titre observed in the second replicate (R2) of the lower MOI BUNV experiment, which was higher than the peak observed in the corresponding replicate of the higher MOI experiment (see Fig~\ref{fig:posterior_predictions_infection}, panels C1 and C2). A potential mechanism that could explain this behaviour of a lower viral titre at high MOI is the presence of defective interfering particles (DIPs) within the virus stock, since the higher probability of co-infection with DIPs at a higher MOI could cause fewer cells to produce standard virus, making the overall viral production rate lower~\cite{liao2016validating}. Therefore, in order to capture this with the MM, we could have incorporated different production rate parameters for each MOI, as done in previous MMs for
influenza virus~\cite{paradis2015impact}.
However, the opposite pattern was observed in the BUNV data for the first replicate (R1 in 
Fig~\ref{fig:posterior_predictions_infection}, panels C1 and C2), where the peak viral titre was actually lower for the lower MOI than the higher MOI, implying that DIPs were not present at an appreciable concentration in the inoculum. A similar pattern of a higher MOI leading to more virus produced has been observed for RSV, implying that only the RSV inoculum caused infection and RSV progeny resulted in a negligible number of infection events~\cite{beauchemin2019uncovering}.
We believe that the differences observed between our experimental replicates are due to random effects and therefore we decided to assume the same viral production rate for both MOIs. However, the observation that BATV also exhibited a slightly higher viral titre plateau at lower MOI (see Fig~\ref{fig:posterior_predictions_infection}, panels C3 and C4) suggests that MOI-dependent effects may be more complicated than captured here.

We used A549 cells for the experiments because they are a well-established system that supports robust infection with both BUNV and BATV, and they have been widely used in experimental studies of orthobunyaviruses~\cite{hover2016modulation, shi2016bunyamwera, bowen2025probing}.
However, A549 cells are derived from human lung epithelium, whereas in natural infections of mammalian hosts, BUNV and BATV are transmitted via arthropod bite, with virus initially entering the skin dermis before disseminating into the bloodstream. The first cells to encounter incoming virus are likely to include resident dendritic cells, macrophages, and other innate immune cells recruited to the site of virus transmission~\cite{leger2015bunyaviruses}.
Therefore, while A549 cells are useful for quantifying viral life cycle kinetics under \textit{in vitro} conditions, they are not the most relevant cell type for extrapolating the results to within-host infection.
Future work could explore alternative cell types that are more physiologically relevant than A549 cells, such as fibroblasts. Macrophages would also be relevant for understanding early infection events, but their use may present challenges, for example due to their suspended nature. Also, since their activation state can change during infection, it may be necessary to incorporate this as an additional biological mechanism in the MM.

In conclusion, this work provides the first quantitative comparison of the \textit{in vitro} infection dynamics of BUNV and BATV, and establishes a foundation for future research on othobunyavirus replication and reassortment dynamics. Applying similar modelling strategies to reassortant strains may help clarify how parental viral traits shape the behaviour of emergent viruses. Finally, incorporating additional experimental data, for example to enable exploration of alternative transmission mechanisms such as cell-to-cell spread, would be an important next step toward a more complete understanding of this \textit{in vitro} system.

\section*{Methods}

\subsection*{Viral decay experiment}

Three different aliquots of BUNV or BATV virus stock were thawed and prepared into three replicates as separate viral master mixes in 2.5\% DMEM media resulting in an infectious virus concentration of approximately $\unit{10^5}{SIN/\milli\litre}$ for BUNV and $\unit{4\times 10^5}{SIN/\milli\litre}$ for BATV. Out of each master mix, \unit{750}{\micro\litre} was placed in each of six wells, one for each of the daily time points \unit{[0,5]}{\dpi}. All three plates were incubated at $\unit{37}{\degree\Celsius}$. At each time point (every \unit{24}{h}), one well from each of the 3 replicates was harvested and stored at $\unit{-80}{\degree\Celsius}$ until titration via endpoint dilution (ED) assay with A549 cells. Prior to their use in the ED assay, the sampled tubes were weighted to give an approximation of the volume loss over their respective incubation time, assuming $\unit{1}{\milli\litre}\approx\unit{1}{\gram}$.

\subsection*{Replication in A549 cells}

A549 cells (adenocarcinomic human alveolar
basal epithelial cells) were seeded in 6-well plates and incubated at $37\degree\Celsius$ overnight. On the day of infection, growth media (10\% FBS DMEM) was removed and cells were washed twice with PBS to remove residual serum. Cells were infected with BUNV or BATV at a low MOI or a high MOI,
or mock infected in 2.5\% FBS DMEM, at a total volume of \unit{2}{\milli\litre}. The supernatant viral titres at the 0 hpi time point indicate that the low MOI was approximately
0.1 SIN/cell and 
0.3 SIN/cell for BUNV and BATV, respectively, and the high MOI was approximately
4.5 SIN/cell and 
24.5 SIN/cell for BUNV and BATV, respectively, in terms of the amount of virus added to the well.
Cells were then incubated for \unit{1}{h} at $37\degree\Celsius$ to allow virus entry into cells. At \unit{1}{h} post infection (hpi), supernatant was removed and the 1 hpi pre-wash sample from one well was stored until titration to obtain an estimate of how much virus had entered cells during the \unit{1}{h} incubation period. After the supernatant was removed at 1 hpi, cells were washed once with 1$\times$ PBS, once with an acid wash (PBS pH 3.0), and a final PBS wash. Media was replaced with \unit{2}{\milli\litre} fresh 2.5\% FBS DMEM and cells were incubated until a specific time point (different for each well).

At the sample time point for a given well, supernatant was collected in \unit{400}{\micro\litre} aliquots and stored at $-80\degree\Celsius$ until downstream
titration by ED assay (both BUNV and BATV) or RNA analysis (BUNV high MOI only). 

To measure the number of cells in the well, the monolayer was washed once with 1$\times$ PBS, then detached by the addition of \unit{300}{\micro\litre} trypsin for 5 minutes at $37\degree\Celsius$. Trypsin was neutralised with \unit{1.2}{\milli\litre} 10\% FBS DMEM and collected into a sterile \unit{1.5}{\milli\litre} tube. From this, \unit{10}{\micro\litre} of cell suspension was taken to assess cell density using a hemocytometer.

For BUNV, the growth curve experiment was initially replicated twice, with slightly different (but overlapping) sets of sample time points up to 74 hpi. The first replicate (R1) included the following time points (hpi): 0, pre-wash (1 hpi), post-washes (1 hpi), 2.5, 6.5, 9.5, 12, 14, 26, 32, 55.5, 74. The second replicate (R2) included the following time points (hpi): 0, pre-wash (1 hpi), post-washes (1 hpi), 2.5, 5, 6.5, 22, 26, 32, 45.5, 51, 55.5, 69, 74. During the \unit{74}{h} time course, we did not observe a clear decline in number of cells or viral titre (even though in the viral decay assay, infectious virus was observed to decay). This indicated that the cells could still be producing virus at \unit{74}{h} and we could not identify the infectious period. Therefore a second experiment was conducted to identify what occurs after \unit{74}{h}. Specifically, this experiment (two replicates, R3 and R4) included the following time points (hpi): 0, pre-wash (1 hpi), 72, 96, 120, 144, 168. At these later time points, we did observe a clear decline in the number of cells, so that the infectious period could be estimated. The method used was the same as described above, but due to difficulties in obtaining enough A549 cells and the extra time needed to conduct the titration using A549 cells, the titration by ED assay was conducted using BHK cells instead. This meant that the observed viral titre plateau was lower.

For BATV, the growth curve experiment was conducted three times, with the first set of sample time points up to 12 hpi, the second set of sample time points from 18 hpi to 72 hpi, and the third set of sample time points from 72 hpi to 168 hpi. The first experiment (two replicates at each time point, R1 and R2) included the following time points (hpi): 0, pre-wash (1 hpi), 4, 8, 12. The second experiment (two replicates at each time point, R3 and R4) included the following time points (hpi): 0, pre-wash (1 hpi), 18, 24, 48, 72. The third experiment (one replicate, R5) included the following time points (hpi): 0, pre-wash (1 hpi), 72, 96, 120, 144, 168. For BATV, all titrations by ED assay were conducted using A549 cells.

\subsection*{Quantification of infectious virus (SIN)}

Infectious virus titres were determined by ED assay on A549 cells or BHK cells. Cells were seeded in 96-well plates at 6 × $10^4$ cells per well in \unit{150}{\micro\litre} of 2.5\% DMEM and incubated overnight at 37°C. Virus-containing samples were serially diluted 10-fold in 2.5\% DMEM, from $10^{-1}$ to $10^{-6}$ for the samples from the viral decay experiments, and from $10^0$ to $10^{-8}$ for the samples from the infection experiments. A given inoculum volume of each dilution was added to six replicate wells of cells and incubated for 5 days at 37°C. The inoculum volumes used were \unit{45}{\micro\litre} for BUNV R1 and R2, \unit{50}{\micro\litre} for BUNV R3 and R4 and all BATV samples, and \unit{100}{\micro\litre} for samples from the viral decay experiments.

Cells were fixed by direct addition of formaldehyde to a final concentration of 4\% for 10 min, washed with water, and stained with 2\% crystal violet in 20\% ethanol. Wells were scored for the presence or absence of cytopathic effect, and infectious titres were calculated
in units of specific infections (SIN) using the method described in Ref.~\cite{midSIN}. SIN is an alternative quantisation to the \TCID \ concentration, and an alternative to estimation by the Reed-Muench and Spearman-Karber methods.

\subsection*{Quantification of viral RNA copies}

Culture supernatant and cell lysate samples were submitted to Azenta Life Sciences for viral genome quantification by reverse transcription digital PCR (RT-dPCR). Viral RNA was extracted using the CGT Viral Vector Lysis Kit (Qiagen) and treated with ezDNase (Invitrogen) to remove contaminating DNA. For extra-cellular samples, \unit{200}{\micro\litre} of culture supernatant was used for each RNA extraction. Approximately \unit{400}{\micro\litre} of supernatant was submitted per sample, and two independent RNA extractions were performed to provide sufficient material for analysis. Reverse transcription was performed using the SuperScript IV First-Strand Synthesis System (Invitrogen) according to manufacturer’s protocol, followed by treatment with Thermolabile Exonuclease I (NEB). 

Digital PCR was performed on a QIAcuity One using QIAcuity Nanoplates and custom primer/probe assays according to Azenta standard protocols. Each reaction had a total volume of \unit{15}{\micro\litre}, comprising \unit{1.25}{\micro\litre} of cDNA template, and \unit{14}{\micro\litre} was loaded onto the nanoplate. Data were analysed using the QIAcuity Software Suite. Partition thresholds were defined using no-template controls and applied consistently across all reactions. The software reported absolute RNA copy numbers as copies per \micro\litre \ of PCR reaction mixture. 

To estimate the total number of viral RNA copies in the original infection well, the reported values were multiplied by a factor of (15/1.25) $\times 2 \times 10^3$. This conversion accounts for the \unit{15}{\micro\litre} total PCR reaction volume, the use of \unit{1.25}{\micro\litre} cDNA template per reaction, and the 2 ml total volume of culture supernatant collected from each well. Final values shown in 
Fig~\ref{fig:posterior_predictions_RNA} are total viral genome copies per infection well. 

\subsection*{Mathematical model (MM)}

To numerically simulate the \textit{in vitro} experiments, we use the ODE model in Eq~\eqref{eq:model_ODEs} and shown in Fig~\ref{fig:model_diagram}. This MM is based on that of Quirouette \textit{et al.}~\cite{quirouette2024does}, but has been extended here to describe (i) resource-limited cell division and time-dependent natural cell death, (ii) entry of virions into cells that are already infected, and (iii) that re-infection of a cell can reduce its delay before progeny virion production.

The variable $V_\ssin$ represents the number of SIN (as measured by our ED assay using A549 cells) in the supernatant per well, and the variable $V_\RNA$ represents the number of RNA copies of a given genome segment in the supernatant (assumed to be equal for each segment length, S, M, and L).
The variable $T$ denotes the number of uninfected target cells per well. The population of infected cells is partitioned into eclipse phase cells, $E = \sum_{i=1}^{n_E}E_i$, and infectious phase cells, $I = \sum_{j=1}^{n_I}I_j$. The MM incorporates a multistage representation of the eclipse and infectious phases, so that the time a cell spends in each phase follows an Erlang distribution.

We assume that SIN (and RNA copies) are lost to cell entry at rate $\beta/s$ per cell, where $s$ is the volume of the supernatant in one well. For each unit of SIN that is lost to entry into uninfected target cells, the number of cells that become infected is represented by the parameter $\gamma$~\cite{quirouette2024does}. When cells become infected, they first enter the eclipse phase, which is the period after virus entry but before the release of virions. The mean time spent in the eclipse phase (with $n_E$ stages) is $\tau_E$, and thus, the rate at which cells move through each eclipse stage is $\frac{n_E}{\tau_E}$. When a cell exits the last stage of the eclipse phase, it enters the infectious phase (with ${n_I}$ stages). The mean time spent in the infectious phase is $\tau_I$, and thus, the rate at which cells move through each infectious stage is $\frac{n_I}{\tau_I}$. During the infectious phase, cells are assumed to release SIN and RNA copies (of each segment length) at a constant rate of $p$ SIN per hour per cell and $p_\RNA$ RNA copies per hour per cell, respectively. Cells stop producing virus upon exiting the last infectious stage.
Extra-cellular SIN in the supernatant decays at rate $c$. We do not include decay of RNA copies since the RNA decay rate is assumed to be negligible.

To model cell division, we define an explicit variable for the availability of resources for cell division, $R(t)$, starting at $R(0)=1$. Cells are assumed to use these at rate $\mu/s$ per cell, and the cell division rate, $\lambda R(t)$, is proportional to the amount of resources available. Therefore, the division rate decreases over time as resources are used up. We allow for division of uninfected and eclipse phase cells, but assume that cells stop dividing when they become infectious. When a cell divides, we assume that it produces two daughter cells in the same 
infection state, $T$ or $E_i$, as itself.

Natural cell death is modelled as a time‑dependent process at the population level, which is defined relative to the start of the experiment, rather than the birth time of individual cells. We assume that the time until natural death of the cells initially present in the well at time \unit{0}{h} follows an Erlang distribution with mean $\tau_N$ and $n_N$ stages and that newly divided cells inherit the same time-dependent death rate as their parent cell. Consequently, when a cell divides, the remaining lifetime of its daughter cells depends on the current time in the experiment, rather than how long those cells have existed individually. To implement this in the MM, we define a time-dependent cell death rate for all cells, $\delta_N(t)$, given by the hazard function of an Erlang-distributed random variable.
This implementation allows us to treat natural cell death independently of infection state. Otherwise, if cells were to reset their lifespan upon division, then since we allow for division of uninfected cells and eclipse phase cells but not infectious cells, cells that die naturally would be more likely to come from the infectious compartment than the other compartments. In this case, we would need to explicitly model the age distribution of cells, which would substantially increase the complexity of the MM.
Although our use of a time-dependent death process at the population level is a simplification, it is consistent with the behaviour that we observed in the \textit{in vitro} cell cultures in the absence of infection, where cell populations initially grew and subsequently declined. We hypothesise that cell death is driven by factors such as the accumulation of waste products, which affect the entire cell population rather than individual cells independently. Therefore, since newly divided cells are subject to the same environment as older cells, we assume that their death rate is influenced by the overall state of the culture, rather than by their individual age.

We assume here that the virions can enter cells that are already infected. We also consider that re-infection of a cell by additional virions during the
eclipse phase may reduce the time taken for the cell to complete the eclipse phase and start producing viral progeny;
that is, the MM assumes that every time an eclipse phase cell in stage $i\leq n_E$ is infected by another virion, the time that the cell spends in the eclipse phase will be shortened by skipping ahead through $\kappa$ eclipse stages to stage $i + \kappa$.
Note that we do not allow cells to skip infectious stages, so if the jump is larger than the number of remaining eclipse stages ({\em i.e.}, $i + \kappa > n_E$), the cell will simply enter the first infectious stage.
If $\kappa=0$, then this implies that cells can still become re-infected, but re-infection has no effect on the duration of a cell's eclipse phase. In
Eq~\eqref{eq:model_ODEs} we define $\boldsymbol{1}_A$ as an indicator function, which is equal to $1$ if $A$ is true, and $0$ otherwise.

\begin{equation}\label{eq:model_ODEs}
	\begin{split}
		\ddt{T}&=\lambda R T - \delta_N(t)T -\gamma\frac{\beta}{s} V_\ssin T,\\[2ex]
		\ddt{E_1}&= \lambda R E_1 - \delta_N(t)E_1 + \gamma\frac{\beta}{s} V_\ssin T  -\frac{n_E}{\tau_E}E_1 - \boldsymbol{1}_{\kappa>0}\gamma\frac{\beta}{s} V_\ssin E_1,\\[2ex]
		\ddt{E_i}&=\lambda R E_i - \delta_N(t)E_i + \frac{n_E}{\tau_E}(E_{i-1}-E_i) -\boldsymbol{1}_{\kappa>0}\gamma\frac{\beta}{s} V_\ssin E_i + \boldsymbol{1}_{0<\kappa<i}\gamma\frac{\beta}{s}V_\ssin E_{i-\kappa},\\ & \qquad \qquad \qquad \qquad \qquad \qquad \qquad \qquad \qquad \qquad \qquad \qquad \qquad \qquad i=2,\ldots,n_E,\\[2ex]
		\ddt{I_1}&= - \delta_N(t)I_1 + \frac{n_E}{\tau_E}E_{n_E} - \frac{n_I}{\tau_I} I_1 + \boldsymbol{1}_{\kappa>0}\gamma\frac{\beta}{s}V_\ssin\sum_{i=n_E-(\kappa-1)}^{n_E} E_i,\\[2ex]
        \ddt{I_j}&= - \delta_N(t)I_j + \frac{n_I}{\tau_I}(I_{j-1}-I_j), \qquad j=2,\ldots,n_I,\\[2ex]
		\ddt{V_\ssin}&=p \sum_{j=1}^{n_I}I_j - \left(c + \frac{\beta}{s}\left(T +\sum_{i=1}^{n_E}E_i + \sum_{j=1}^{n_I}I_j\right) \right)V_\ssin,\\[2ex]
        \ddt{V_\RNA}&=p_\RNA \sum_{j=1}^{n_I}I_j - \frac{\beta}{s}\left(T +\sum_{i=1}^{n_E}E_i + \sum_{j=1}^{n_I}I_j\right) V_\RNA,\\[2ex]
        \ddt{R}&= - \frac{\mu}{s} R \left(T +\sum_{i=1}^{n_E}E_i + \sum_{j=1}^{n_I}I_j\right),
	\end{split}
\end{equation}
with $\delta_N(t) = \frac{\left(\frac{n_N}{\tau_N}t\right)^{n_N}}{t(n_N-1)!\displaystyle\sum_{k=0}^{n_N-1}\frac{1}{k!}\left(\frac{n_N}{\tau_N}t\right)^k}$.
\\[3ex]

We note that the supernatant samples from our BUNV experiments focussing on \unit{72}{h} onward were titrated using BHK cells in the ED assay (instead of A549 cells) and the measured number of SIN was noticeably different ({\em e.g.}, due to cell type differences in susceptibility to infection). Therefore, we introduce the parameter $\alphaB$ as the ratio between the number of SIN measured with BHK cells ($V_\ssin^\text{BHK}$) and the number of SIN measured with A549 cells ($V_\ssin$), leading to the following relationship.

\begin{eqnarray*}
    V_\ssin^\text{BHK}(t) & = & \alphaB V_\ssin(t).\\
\end{eqnarray*}

\subsection*{Parameter inference}

Parameter inference for the MM in Eq~\eqref{eq:model_ODEs} and Fig~\ref{fig:model_diagram} was performed using Markov chain Monte Carlo (MCMC) with \texttt{phymcmc}~\cite{phymcmc}, to estimate the posterior distribution of the parameters. The posterior distribution is proportional to the likelihood function multiplied by the prior distribution, which can be written as
\begin{align*}
\Po(\pars|\text{data}) &\propto \Lik(\text{data}|\pars) \cdot \Pri(\pars),
\end{align*}
where $\pars$ is the vector of unknown model parameters.

\begin{figure}[H]
\centering
    \includegraphics[width=\textwidth]{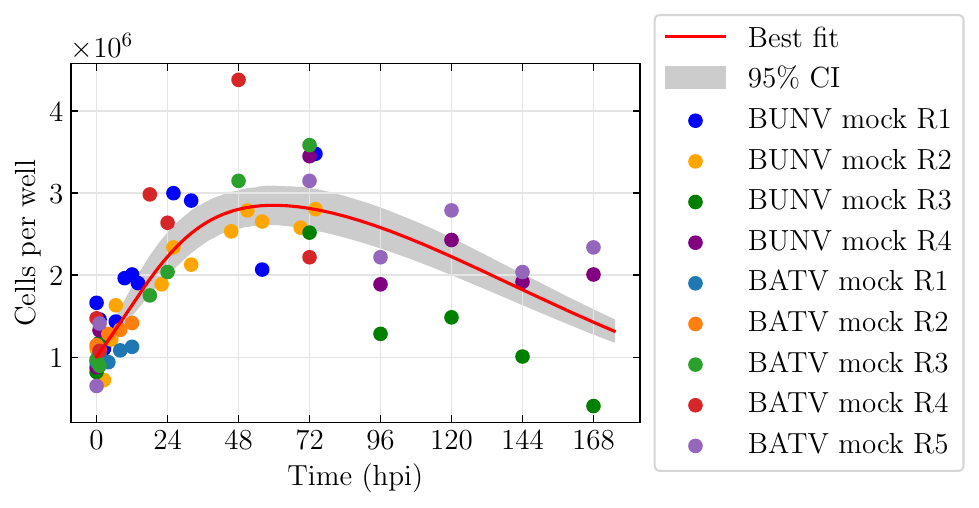}
    \caption{\textbf{Posterior MM predictions for cell division and natural death, compared with the cell count data from the mock infection experiments.}}
    \label{fig:mock_cells_prediction}
\end{figure}

Before inferring the parameters for BUNV and BATV, we estimated the posterior distribution of the model parameters that describe cell division and natural death: $\pars = (T_0, \ \lambda, \ \tau_N, \ n_N, \ \mu)$. This was done using the cell count data from the experiments in which cells were mock infected in 2.5\% DMEM. Fig~\ref{fig:mock_cells_prediction} shows the comparison between the posterior MM predictions for cell division and natural death compared to the observed cell counts from these mock experiments.

We assumed log-uniform prior distributions for the parameters $T_0$, $\lambda$, and $\mu$. We assumed linearly uniform prior distributions for the parameters $\tau_N$ and $n_N$. Therefore, the overall prior distribution is such that
\begin{align*}
\Pri(\pars) &\propto \frac{1}{T_0 \cdot \lambda \cdot \mu}.
\end{align*}

The likelihood function, $\Lik(\text{data}|\pars)$, describes the likelihood of the observed experimental data, given the MM prediction corresponding to a set of parameter values, $\pars$. To define the likelihood of the measured cell numbers, given the MM and a set of parameters $\pars$, we made the assumption that the residuals between the logarithms of the experimentally-measured and MM-predicted number of cells follow a normal distribution. This gives a log-likelihood of
\begin{align*}
\ln \Lik_\text{cells}(\text{data}|\pars) &= \sum_{t, r} \ -\frac{\left(\ln(C_{t, r}^\text{data})-\ln(N(t\mid \pars))\right)^2}{2\sigma_C^2},
\end{align*}
where $C_{t, r}^\text{data}$ is the experimentally-measured number of cells per well at time $t$ and replicate $r$, and $N(t|\pars)=T(t)$ is the MM-predicted number of cells at time $t$ in the mock infection experiment, obtained from the system of 
equations~\eqref{eq:model_ODEs} with initial conditions $T(0)=T_0$ (estimated), $E_i(0)=I_j(0)=0$ for $i = 1, \ldots , n_E$ and $j = 1, \ldots , n_I$, $V_\ssin(0)= 0$, $V_\RNA(0)= 0$, and $R(0)=1$. The standard deviation ($\sigma_C = 0.28$ for the BUNV mock experiments and $\sigma_C = 0.19$ for the BATV mock experiments), was estimated by calculating the standard deviation of the residuals between the $\ln(C_{t, r}^\text{data})$ measurements at each sampling time and their mean, pooled over all time points and replicates.

In the subsequent parameter inference for BUNV and BATV, we fixed the values of $\lambda$, $\tau_N$, $n_N$, and $\mu$ to their most likely values given by their multi-dimensional posterior distribution 
(see Table~\ref{table:parameters} for the fixed values). Then, for each virus separately, the posterior distribution for the parameter vector, $\pars = (T_0, \beta, c, \gamma, p, p_\RNA, n_E, n_I, \kappa, \tau_E, \tau_I, \omega, \alphaB)$ was estimated by simultaneously using the \textit{in vitro} data from the viral decay experiment and the infection experiments with A549s. From the viral decay experiment, we have viral titre measurements over time, and from the infection experiments we have viral titres and cell numbers over time. From the BUNV higher MOI experiment we additionally have information on the numbers of negative-sense RNA copies of each segment in the supernatant over time.

When using the MM in \eqref{eq:model_ODEs} to simulate the viral decay experiment, we used an initial condition of $V_\ssin(0)= V_0^\text{decay}$, where $V_0^\text{decay}$ is the geometric mean of the most likely values (modes) of SIN/well calculated by 
midSIN~\cite{midSIN} using the ED assay outcomes of the three replicates at time \unit{0}{\hour}. Thus, since there were no cells present in this experiment ($T(t)=E_i(t)=I_j(t)=0$ for all times), the MM prediction for the number of SIN in a well at time $t$ is
$$V_\ssin(t) = V_0^\text{decay}e^{-c t}.$$

When using the MM in \eqref{eq:model_ODEs} to simulate the infection experiment, we used initial conditions of $T(0)= T_0$ (estimated), $E_i(0)=I_j(0)=0$ for $i = 1, \ldots, n_E$ and $j = 1, \ldots, n_I$, $V_\ssin(0)= V_0^\text{MOI}$ ($\text{MOI}\in\{\text{low}, \ \text{high}\}$), and $R(0)=1$. The initial number of SIN for each MOI experiment ($V_0^\text{MOI}$) was set to the corresponding geometric mean of the most likely values (modes) calculated by midSIN using the ED assay outcomes for the $0$ hpi time point. We only considered the variable $V_\RNA$ for the high MOI BUNV experiment, in which case the initial condition, $V_\RNA(0) = V_0^\RNA$, was set to the geometric mean of the number of RNA copies of each genome segment at the $0$ hpi time point.
We note that the wash steps conducted after the \unit{1}{\hour} incubation period are unlikely to be 100\% effective in removing all extra-cellular virus. Therefore, we introduced the parameter $\omega\in [0, 1]$ to represent the fraction of extra-cellular virus remaining after the wash steps ($\omega = 0$ implies a perfect wash, whereas $\omega = 1$ implies no wash). We simulated the wash steps by setting
\begin{align*}
    V_\ssin(1 \textrm{ hpi})_{\textrm{post-wash}} &= \omega V_\ssin(1 \textrm{ hpi})_{\textrm{pre-wash}},\\
    V_\RNA(1 \textrm{ hpi})_{\textrm{post-wash}} &= \omega V_\RNA(1 \textrm{ hpi})_{\textrm{pre-wash}},
\end{align*}
where for example$V_\ssin(1 \textrm{ hpi})_{\textrm{pre-wash}}$ and $V_\ssin(1 \textrm{ hpi})_{\textrm{post-wash}}$ are the number of extra-cellular SIN at $1$ hpi, immediately prior to and after the wash, respectively. For BATV, the wash factor was fixed to the median value from the BUNV parameter inference ($\omega = 8\times 10^{-4}$), since it was not possible to estimate this parameter from the BATV infection data, due to not having enough early data points to observe a dip in viral titre after the wash and before virus starts to be produced. The ratio between SIN measured in BHK cells and SIN measured in A549 cells ($\alphaB$) was also not estimated for BATV since none of the titrations were conducted using BHK cells, and neither was the production rate of RNA ($p_\RNA$) since we do not have data on RNA copies for BATV.

We assumed log-uniform prior distributions for the parameters $T_0$, $\beta$, $c$, $\gamma$, $p$, $p_\RNA$, $\omega$, and $\alphaB$. We assumed linearly uniform prior distributions for all other parameters.
In addition to the prior bounds in
Table~\ref{table:parameters}, for BUNV we imposed a constraint on $\gamma$ by rejecting (assigning a prior probability of zero to) any value of $\gamma$ that results in a predicted maximum number of cells able to become infected by the initial inoculum ($\gamma V_0^\text{high}$) exceeding the predicted number of viral RNA copies at that time ($V_0^\RNA$) in the high MOI experiment. This resulted in an upper bound of $\gamma\leq 8$.
We also imposed $\gamma p \leq p_\RNA$, but this did not have an impact because the constraint that involved the initial conditions was more stringent.

Due to the high degree of correlation between $\beta$ and $\gamma$, we instead sampled the parameter transformations $a=\beta \cdot \gamma$ and $b= \gamma/\beta$ during the MCMC algorithm~\cite{quirouette2024does}. Given independent log-uniform distributions for $\beta$ and $\gamma$, it can be shown that the joint probability density function of $a$ and $b$ is proportional to $1/(ab)$. Therefore, the overall prior distribution is such that
\begin{align*}
\Pri(\pars) &\propto \frac{1}{a\cdot b\cdot T_0\cdot c\cdot p\cdot p_\RNA \cdot \omega \cdot\alphaB}.
\end{align*}

The likelihood function that we used consists of three parts based on the different types of observed experimental data: (i) ED assay outcomes from the viral decay experiment and the infection experiments, (ii) cell numbers from the infection experiments, and (iii) negative-sense RNA copy numbers from the BUNV high MOI infection experiments.

To define the likelihood of the observed ED assay outcomes, given the MM and a set of parameters $\pars$, we follow an approach that incorporates all the information provided by the ED assay~\cite{quirouette2024does}. Here, for time point, $t$, and replicate, $r$, the total number of replicate wells for dilution factor $d$ in the ED assay is indicated by $n_{t, r, d}$ and the number of these wells that were observed to become infected is $k_{t, r, d}$.
For time point $t$, we assume that the number of SIN in the total sampled supernatant of the experiment well is $V_\text{SIN}(t\mid \pars)$, which is obtained from the MM solution for either the viral decay or infection experiment simulation. Then, the number of SIN inoculated into an ED assay well is assumed to be Poisson distributed with mean $V_\text{SIN}(t\mid \pars)\cdot f_{t, r, d}$, where $f_{t, r, d}$ is the fraction of total supernatant volume inoculated into each of the ED assay wells for dilution factor $d$. Thus, the MM-predicted probability that at least one SIN is inoculated into a given well (and hence infection is established in the well) is
$$q_{t, r, d}(\pars) = 1-\exp\left[-V_\text{SIN}(t\mid \pars)\cdot f_{t, r, d}\right].$$
Therefore, for a given time point, $t$, and replicate, $r$, the overall likelihood of the observed ED assay outcome ($k_{t, r, d}$ out of $n_{t, r, d}$ wells observed to be positive for infection at each dilution, $d$) is proportional to
$$\prod_{d} q_{t, r, d}(\pars)^{k_{t, r, d}}\cdot (1-q_{t, r, d}(\pars))^{(n_{t, r, d} - k_{t, r, d})}.$$
We then take the product over all replicates and time points (across both the viral decay experiment and infection experiment) to give a log-likelihood of
\begin{align}\label{eqn:lik_ED}
\ln \Lik_\text{ED}(\text{data}|\pars) &= \sum_{t,r,d} \ k_{t,r,d}\cdot \ln(q_{t,r,d}(\pars)) + (n_{t,r,d} - k_{t,r,d})\cdot \ln(1- q_{t,r,d}(\pars)).
\end{align}

For the likelihood of the measured cell numbers, given the MM and a set of parameters $\pars$, we made the assumption that the residuals between the logarithms of the experimentally-measured and MM-predicted number of cells follow a normal distribution. This gives a log-likelihood of
\begin{align*}
\ln \Lik_\text{cells}(\text{data}|\pars) &= \sum_{t, r} \ -\frac{\left(\ln(C_{t, r}^\text{data})-\ln(N(t\mid \pars))\right)^2}{2\sigma_C^2},
\end{align*}
where $C_{t, r}^\text{data}$ is the experimentally-measured number of cells per well at time $t$ and replicate $r$, $N(t|\pars)=T(t) +\sum_{i=1}^{n_E}E_i(t) + \sum_{j=1}^{n_I}I_j(t)$ is the MM-predicted number of cells at time $t$, and $\sigma_C$ is the estimated standard deviation of the residuals. The standard deviation was estimated separately for each virus and MOI by calculating the standard deviation of the residuals between the $\ln(C_{t, r}^\text{data})$ measurements at each sampling time and their mean, pooled over all time points and replicates. For BUNV low MOI, $\sigma_C = 0.23$, and for BUNV high MOI, $\sigma_C = 0.26$. For BATV low MOI, $\sigma_C = 0.28$, and for BATV high MOI, $\sigma_C = 0.26$.

For the likelihood of the measured number of extra-cellular genome copies of each segment, given the MM and a set of parameters $\pars$, we made the assumption that the residuals between the logarithms of the experimentally-measured and MM-predicted number of genome copies follow a normal distribution. We have

\begin{align*}
\ln \Lik_\text{RNA}(\text{data}|\pars) &= \sum_{t, r, l} \ -\frac{\left(\ln(R_{t, r, l}^\text{data})-\ln(V_\RNA(t\mid \pars))\right)^2}{2\sigma_{R,l}^2},
\end{align*}
where $R_{t, r, l}^\text{data}$ is the experimentally-measured number of RNA copies of segment $l\in\{S, M, L\}$ at time $t$ and replicate $r$, 
$V_\RNA(t|\pars)$ is the MM-predicted number of genome copies at time $t$, and $\sigma_{R, l}$ is the estimated standard deviation of the experimental measurement error for segment $l$. The standard deviation was estimated separately for each segment length, $l$, by calculating the standard deviation of the residuals between the $\ln(R_{t, r, l}^\text{data})$ measurements at each sampling time and their mean, pooled over all time points and replicates. The values are $\sigma_{R, S} = 0.76$, $\sigma_{R, M} = 1.05$, $\sigma_{R, L} = 0.49$.

The overall log-likelihood function is then
\begin{equation*}
\ln \Lik(\text{data}|\pars) = \ln \Lik_\text{ED}(\text{data}|\pars) + \ln \Lik_\text{cells}(\text{data}|\pars) + \ln \Lik_\text{RNA}(\text{data}|\pars).
\end{equation*}

\section*{Data availability statement}

Computer codes (in Python) for reproducing the results in Figs 3 - 9 are available from the GitHub repository: https://github.com/Bevelynn/BUNV-BATV-mathematical-model.

\section*{Funding}

This research has been supported by the UK Research and Innovation (UKRI, \url{https://www.ukri.org/}) Biotechnology and Biological Sciences Research Council (BBSRC, \url{https://www.ukri.org/councils/bbsrc/}) through project reference BB/W010755/1 (BW, ET, ABS, JNB, GL, and MLG).
Research reported in this publication
was supported by  the
National Institute of Allergy and Infectious Diseases
 of the National Institutes of Health under award numbers
R01AI167048 to CMP.
 This work was also supported in part by Discovery Grant 2022-03744 (CAAB)
from the Natural Sciences and Engineering Research Council of Canada
(\url{https://nserc-crsng.canada.ca/}), and by the RIKEN Center for
Interdisciplinary Theoretical and Mathematical Sciences (iTHEMS,
\url{ithems.riken.jp}) (CAAB).
This collaboration and research was made possible by a Research-in-Groups programme funded by the International Centre for Mathematical Sciences, Edinburgh.
The content is solely the responsibility of the authors and does not necessarily represent the official views of the National Institutes of Health.
The funders had no role in study design, data collection and analysis, decision to publish, or preparation of the manuscript.

\section*{Acknowledgments}
This manuscript has been internally reviewed at Los Alamos National Laboratory, and assigned the reference no. LA-UR-26-24934.

\nolinenumbers

\end{document}